\documentclass{aa}
\usepackage[dvips,final]{graphics}
\usepackage{natbib}
\usepackage{amsfonts}
\usepackage{amsmath}
\usepackage{amssymb}
\usepackage{amstext}

\begin{document}


   \title{Constraining the Cosmological Parameters using Strong Lensing}

   \author{Ghislain Golse
          \and
           Jean-Paul Kneib
          \and
           Genevi\`eve Soucail          }

   \offprints{Ghislain Golse, \email{ghislain.golse@ast.obs-mip.fr}}

   \institute{Laboratoire d'Astrophysique, Observatoire Midi-Pyr\'en\'ees,
              14 av. E.-Belin, F-31400 Toulouse, France
             }

   \date{Received 2001 / Accepted 2001}

   \abstract{ We investigate the potentiality of using strong lensing
     clusters to constrain the cosmological parameters $\Omega_m$ and
     $\Omega_\lambda$.  The existence of a multiple image system with
     known redshift allows, for a given ($\Omega_m$, $\Omega_\lambda$)
     cosmology, absolute calibration of the total mass deduced from
     lens modelling. Recent {\it Hubble Space Telescope} ({\it HST})
     observations of galaxy clusters reveal a large number of multiple
     images, which are predicted to be at different redshifts. If it
     is possible to measure spectroscopically the redshifts of {\it
       many} multiple images then one can in principle constrain
     ($\Omega_m$, $\Omega_\lambda$) through ratios of angular diameter
     distances, {\em independently} of any external assumptions.  For
     a regular/relaxed cluster observed by {\it HST} with 3 multiple
     image systems, each with different spectroscopic redshifts, we
     show by analytic calculation that the following uncertainties can
     be expected: $\Omega_m=0.30 \ \pm 0.11$, $\Omega_\lambda=0.70 \ \pm
     0.23$ or $\Omega_m=1.00 \ \pm 0.17$, $\Omega_\lambda=0.00 \ \pm 0.48$
     for the two most popular world models.  Numerical tests on
     simulated data confirm these good constraints, even in the case
     of more realistic cluster potentials, such as bimodal clusters,
     or when including perturbations by galaxies.  To investigate the
     sensitivity of the method to different mass profiles, we also
     use an analytic ``pseudo-elliptical'' Navarro, Frenk \&
     White profile in the simulations. These constraints can be
     improved if more than 3 multiple images with spectroscopic
     redshifts are observed, or by combining the results from
     different clusters. Some prospects on the determination of the
     cosmological parameters with gravitational lensing are
     given.  
      }

\authorrunning{Golse et al.}
\titlerunning{Constraining the Cosmological Parameters using Strong
 Lensing}
\maketitle

\section{Introduction}
\begin{figure}
  \resizebox{\hsize}{!}{\includegraphics{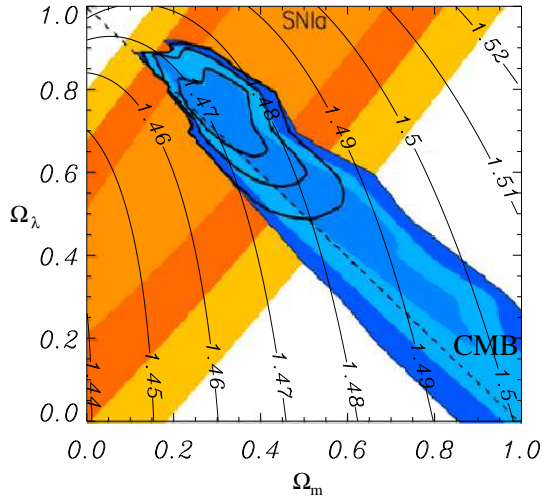}}
\caption{
  Constraints on $(\Omega_m, \Omega_\lambda)$ derived from the most
  recent results of the BOOMERANG and MAXIMA-I experiments on the CMB
  fluctuations and the last results from the SNIa analysis (from
  Jaffe, 2001). Overplotted is the ratio of the geometrical factor of
  the lens equation for two source redshifts, $e(z_{S1}, z_{S2}) =
  E(z_{S1})/ E(z_{S2})$ as discussed in the text. In this example, the
  redshifts are chosen as representative of a typical lens
  configuration: $z_\mathrm{L} = 0.3$, $z_{S1} = 0.7$ and $z_{S2} = 2$.}
\label{E2omla}
\end{figure}

A new ``standard cosmological model'' has arisen in the last few years,
favoring a flat Universe 
with $\Omega_m\sim 0.3$ and $\Omega_\lambda\sim
0.7$. This is mainly based on the results of two experiments which 
give roughly orthogonal constraints in the $(\Omega_m,\Omega_\lambda)$
plane (see Fig.~\ref{E2omla} for a recent update). 
The first one is obtained by considering type Ia supernovae
(SNIa) as standard candles.
The detection of a sample of high redshift SNIa (up to $z \sim$ 1) by
two groups favours a non-vanishing cosmological constant
\citep{Perlmutter,Riess}, large enough to produce an accelerating
expansion. However, evidence for a non-zero cosmological constant is
still controversial, since supernovae might evolve with redshift
and/or may be dimmed by intergalactic dust \citep{Aguirre}. The
fundamental assumption of a homogeneous Universe and its implication
for a non-zero cosmological constant are also discussed
\citep{Celerier, Kolatt}.  The second constraint is derived from the
location of features in the cosmic microwave background (CMB)
anisotropy spectrum, particularly the first Doppler peak. The most
recent results obtained with the Boomerang and MAXIMA experiments
favor a flat Universe \citep{Balbi,
  Melchiorri}. However, there still remains a degeneracy in the
combination of $\Omega_m$ and $\Omega_\lambda$ because CMB experiments
are primarily sensitive to the total curvature of the Universe. Even with the
accuracy of the future {\it MAP} and {Planck} missions, the constraint
issued from the CMB alone will be degenerate.

The combination of these two sets of constraints has led to the
currently favored model of low matter density and a non-zero
cosmological constant, preserving a flat geometry \citep[e.g. ][]{White,
  Efstathiou, Freedman, Sahni, Jaffe}. Although these recent results
are quite spectacular, there still remain many sources of
uncertainties with both methods. Thus any other 
independent test to constrain the large scale 
geometry of the Universe is important to investigate.
Gravitational lensing, an effect involving large distance
scales, has been considered as a very promising tool
for such determinations. Indeed, the statistics of gravitational
lenses depend on the cosmological parameters via angular size
distances and the comoving spherical volume \citep[e.g. ][]{Turner1, Turner2,
  Kochanek, Falco}.  This technique has provided an upper limit on
$\Omega_\lambda$ using different surveys of galaxy lens systems: multiple
quasar statistics \citep{Kochanek,Chiba}, lensed radio sources
\citep{Cooray}, lensed galaxies in the Hubble Deep Field
\citep{Coorayal}. Although most authors favor a lambda-dominated flat
Universe, there remain some uncertainties in the mass distribution of
the galaxy lenses and on the luminosity function of the sources.
Evolutionary effects may also play a role in these statistics.

Another application of gravitational lensing to constrain the cosmological
parameters is to use the statistics of the ``cosmic'' shear variance.
\citet{Waerbeke} showed that it is related to the
power spectrum of the large scale mass fluctuations, and then to
$\Omega_m$. The first results of deep wide field imaging surveys favor
$\Omega_m$ in the range [0.2--0.5] \citep{Maoli,Waerbeke2,Bacon,Wittman}. 
Imaging surveys with the next generation of panoramic CCD cameras will
reinforce this very promising technique. In the case of weak lensing
by clusters of galaxies, \citet{Lombardi} and \citet{Gautret} suggested 
methods to constrain the geometry of the Universe.
These methods need however to recover the mass distribution and/or to know
acurately the redshift of a huge number of distant galaxies, making this method
not practical in the near future.

In this paper, we focus on a measurement technique of
$(\Omega_m,\Omega_\lambda)$ using gravitational lensing as {\sl a
  purely geometrical test of the curvature of the Universe}, since the
lens equation depends on the ratio of angular size distances which is
sensitive to the cosmological parameters. In the most favorable case,
a massive cluster of galaxies can lens several background
galaxies, splitting the images into several families of multiple images.
The existence of one family of multiple images, at known redshift,
allows to calibrate the total cluster mass in an absolute way. In the
case of several sets of multiple images with known redshifts, it is
possible in principle to constrain the geometry of the Universe. This
method was pointed out by \citet{Blandford}, and earlier suggested by
\citet{Paczynski}, but the uncertainties in any lens studies were
considered too large compared to the small variations induced by the
cosmological parameters.  More recently, \citet{Link} (hereafter LP98)
re-analysed the question in the light of the typical accuracy reached
with HST images of clusters of galaxies. Following their method, which
inspired our work, we try to quantify in this paper what can be
reasonably obtained on $(\Omega_m,\Omega_\lambda)$ from accurate lens
modeling of realistic cluster-lenses.

The paper is organized as follows. In Sect. 2 we summarize the main
lensing equations and we introduce the relevant angular size distance
ratio which contains the dependence on the cosmological parameters.
The variation of this ratio is then compared to that of other variables (lens
potential parameters and redshifts) to derive the expected
uncertainties on $\Omega_m$ and $\Omega_\lambda$. In Sect. 3 we
present the method in detail and the results from simulations of
various types of families of images and of different types of lens
potentials. Some conclusions and prospects for the application to real
data are discussed in Sect. 4. Throughout this paper we assume $H_0=65$ 
km s$^{-1}$ Mpc$^{-1}$ (note however that the proposed method and results are 
independent of $H_0$).

\section{Influence of $\Omega_m$ and $\Omega_\lambda$ on image 
formation}
\subsection{Cosmology dependent lensing relations}

We first define the basic mathematical framework, following the
formalism presented in \citet{Schneider}. We consider a lens at a
redshift $z_\mathrm{L}$ with a two-dimensional projected mass distribution
$\Sigma(\vec{\theta})$ and a projected gravitational potential
$\phi(\vec{\theta})$, where $\vec{\theta}$ is a two-dimensional vector
representing the angular position. A source galaxy with redshift $z_\mathrm{S}$
is located at position $\vec{\theta_S}$ in the absence of a lens, and
its image is at position $\vec{\theta_{I}}$. In the lens equation

\begin{equation}
\left\lbrace
\begin{array}{l}
\vec{\theta_S} =\vec{\theta_{I}}-\vec{\nabla} \varphi(\vec{\theta_{I}})\\
\varphi(\vec\theta)=\displaystyle{\frac{2}{c^{2}}}
\displaystyle{\frac{D_\mathrm{LS}}{D_\mathrm{OL}D_\mathrm{OS}}}\phi
(\vec\theta),
\end{array}
\right.
\end{equation}

\noindent $D_\mathrm{OL}$, $D_\mathrm{LS}$ and $D_\mathrm{OS}$ are 
respectively the angular diameter
distances from the Observer to the Lens, from the Lens to the Source
and from the Observer to the Source \citep{Peebles}. $\varphi$ is the reduced
gravitational potential which satisfies $\nabla^2 \varphi=2 \ \Sigma /
\Sigma_{\rm crit}$ with the critical density

\begin{equation}
\Sigma_{\rm crit}=\displaystyle{\frac{c^{2}}{4\pi
    G}}\displaystyle{\frac{D_\mathrm{OS}}{D_\mathrm{LS}D_\mathrm{OL}}}.
\label{crit}
\end{equation}

In these equations the dependence on the cosmological parameters
appears only through the angular diameter distances ratios
$F=\displaystyle{\frac{D_\mathrm{OL} \ D_\mathrm{LS}}{D_\mathrm{OS}}}$ and
$E=\displaystyle{\frac{D_\mathrm{LS}}{D_\mathrm{OS}}}$ ($E$ as {\it
  ``efficiency''}) for a fixed cluster redshift.  They correspond to a
scaling of the lens equation, reflecting the geometrical properties of
the Universe.

In the general case, we can scale the potential gradient as: 
\begin{equation}
\vec\nabla \phi(\vec{\theta_I})=
\sigma_0^2 \ D_\mathrm{OL} \ \vec f(\vec{\theta_I},\theta_C,\alpha,
\ldots)
\end{equation}

where $\sigma_0$ is the central velocity dispersion and $\vec f$ a
dimensionless function that describes the mass distribution of the
cluster. It can be represented by fiducial parameters such as a core
radius $\theta_C$ or a mass profile gradient $\alpha$. The lens
equation reads
\begin{eqnarray}
\label{Eq_Ef}
\vec{\theta_S} & = & \vec{\theta_{I}}-\frac{\sigma_0^2}{c^2} \ 
\frac{D_\mathrm{LS}}{D_\mathrm{OS}} \ 
\vec f(\vec{\theta_{I}},\theta_{C},\alpha, \ldots ) \\
 & = & \vec{\theta_{I}} - \frac{\sigma_0^2}{c^2} \ 
\vec f(\vec{\theta_{I}},\theta_{C},\alpha, \ldots ) \times 
E(\Omega_m, \Omega_\lambda , z_\mathrm{L}, z_\mathrm{S} ) \nonumber
\end{eqnarray}
 
We will focus on the $E$-term which entirely contains the dependence
on $\Omega_m$ and $\Omega_\lambda$ and which is {\sl independent} of
$H_0$.

\subsection{The E-term}
\begin{figure*}
  \hspace{1.0cm}
  \resizebox{\hsize}{!}{
\includegraphics{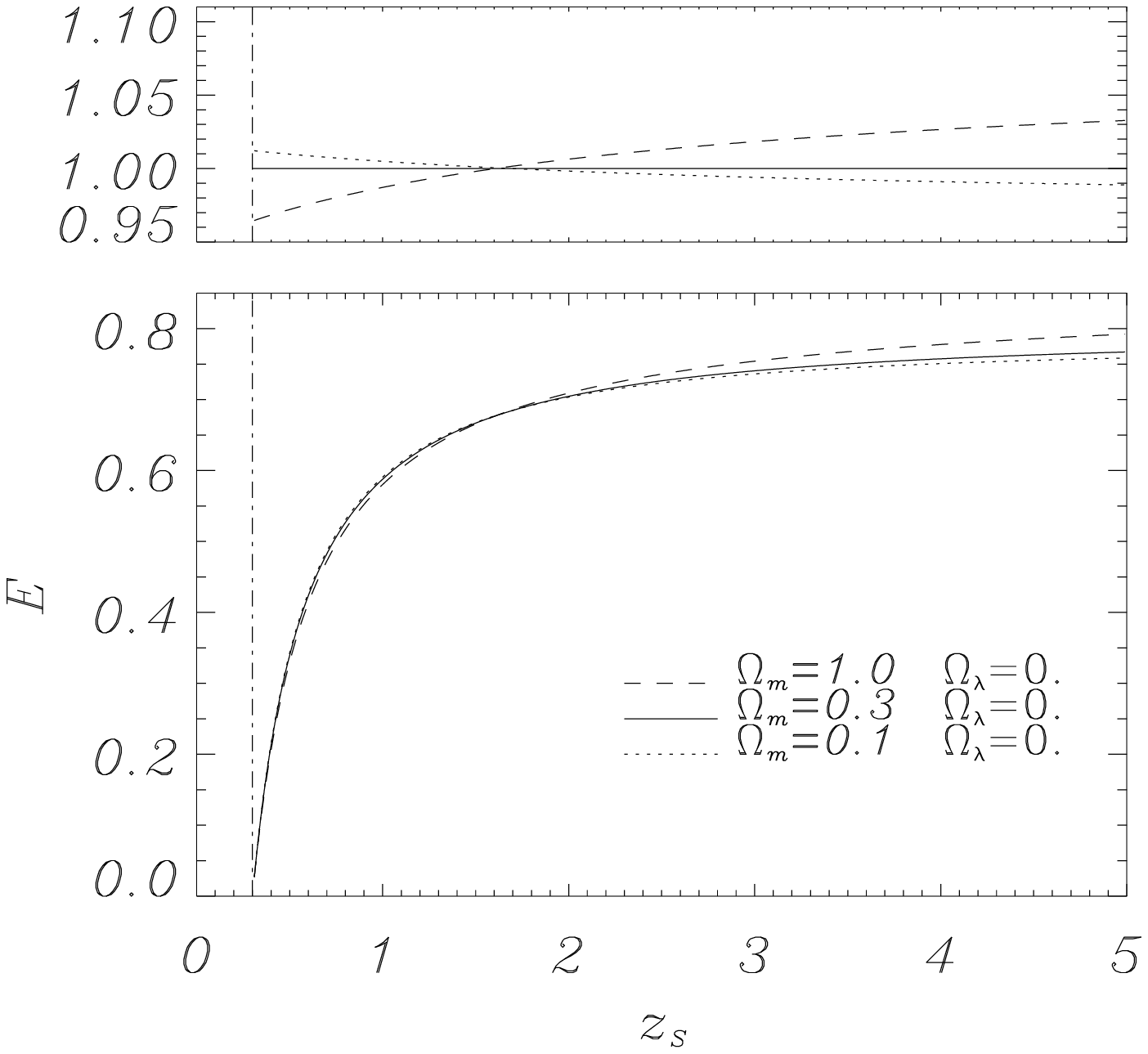}
\includegraphics{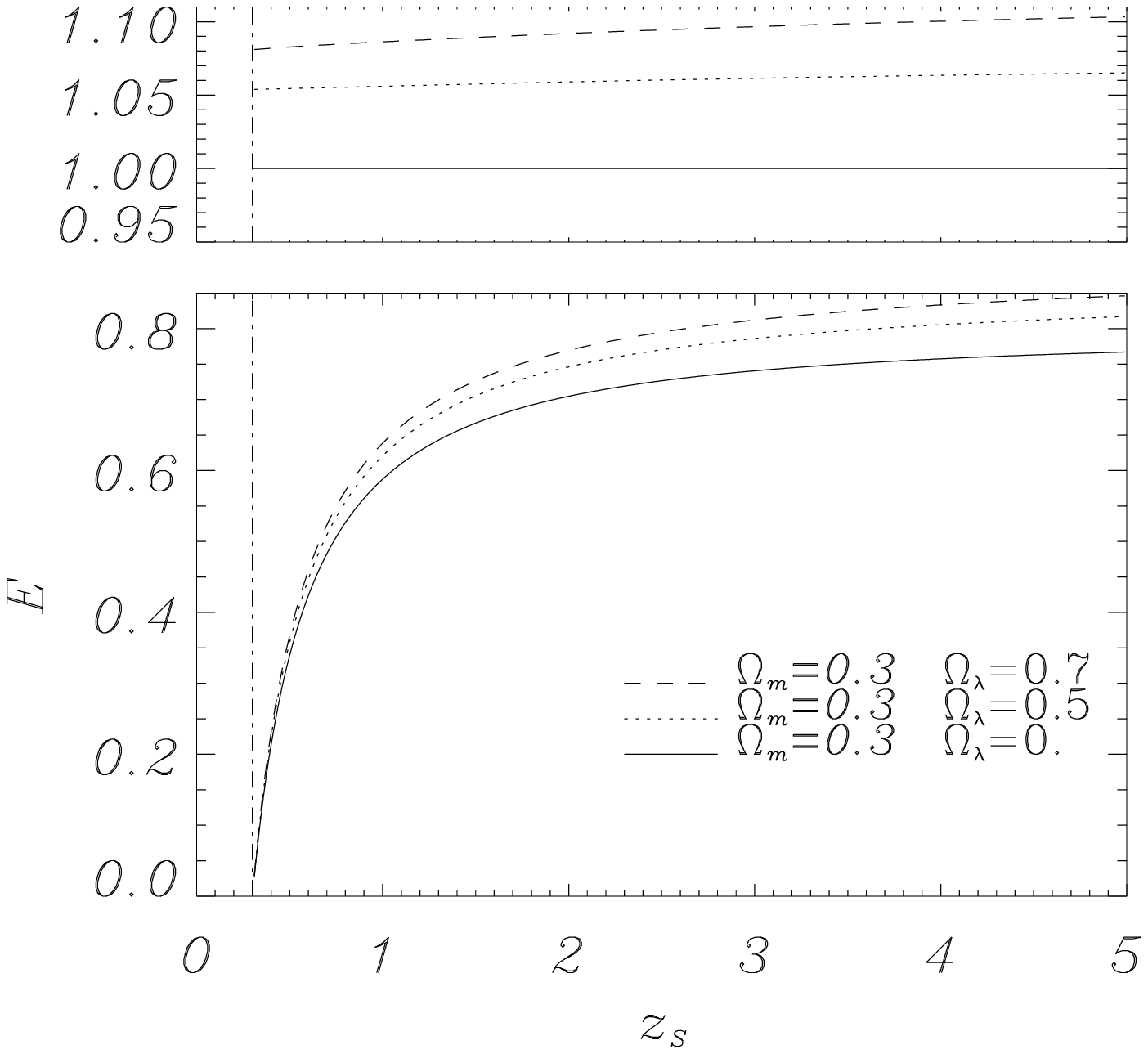}}
\caption{
  Variation of the lens efficiency $E(z_\mathrm{S})$ as a function of
  the source redshift
  for different sets of cosmological parameters:. {\bf (Left):}
  $\Omega_\lambda = 0$ and $\Omega_m$ varies from 0.1 to 1.0. {\bf
    (Right):} $\Omega_m = 0.3$ and $\Omega_\lambda$ varies from 0 to
  0.7. In both cases, the lens redshift is $z_\mathrm{L}=0.3$. For clarity,
  above each main plot, we also plotted the same curves, normalised with
  the $E(z_\mathrm{S})$ function for $(\Omega_m,\Omega_\lambda)=(0.3,0.)$. }
\label{E_zs}
\end{figure*}

For a given lens plane $z_\mathrm{L}$, the ratio $E$ increases rapidly as the
source redshift increases, and then flattens at $z_\mathrm{S}\sim1.5$.
There are also small but significant changes with the cosmological
parameters (Fig.~\ref{E_zs}). The dependence of $E$ with respect to
$\Omega_m$ is weak, whereas the variation with $\Omega_\lambda$ is
larger.

We now consider fixed redshifts for the lens and sources. Assuming a fixed
world model, a single
family of multiple images can in principle constrain the total cluster
mass as well as the shape of the potential, removing the unknown position
of the source $\vec{\theta_S}$ using Eq.(\ref{Eq_Ef}).
In practice, good constraints on the shape of the potential $\vec f$
are obtained with triple, quadruple or quintuple image systems. However 
the absolute normalization
$\sigma_0$ of the mass is degenerate with the E-term, that is with
respect to $\Omega_m$ and $\Omega_\lambda$.

\subsection{Ratio of E-terms for 2 sets of source redshifts}
\label{Eterms} 
To break this degeneracy a second family of multiple images is needed.
To get rid of the strong dependence on $\sigma_0$, it is useful to
consider the ratio of the positions: 
\begin{equation}
\displaystyle
\frac{\|\vec{\theta_{I_1^1}} - \vec{\theta_{I_2^1}}\|}
     {\|\vec{\theta_{I_1^2}} - \vec{\theta_{I_2^2}}\|} 
= \frac{E(z_{S1}) 
\|\vec f(\vec{\theta_{I^1_1}},\ldots) -\vec f(\vec{\theta_{I^1_2}},\ldots )\|}
{E(z_{S2}) 
\|\vec f(\vec{\theta_{I^2_1}},\ldots) -\vec f(\vec{\theta_{I^2_2}},\ldots )\|}
\end{equation}

\noindent (here and hereafter, the
superscript refers to a family and the subscript to a particular image
within a family). This ratio is plotted in Fig.~\ref{E2omla}, highlighting 
the influence of $\Omega_m$ and $\Omega_\lambda$ 
through the ratio $\displaystyle e(z_{S1},z_{S2})=E(z_{S1})/E(z_{S2})$.

Note that the discrepancy between the different cosmological parameters is not
very large, less than 3\% between the Einstein-de Sitter model (EdS)
and a flat, low matter density one. Moreover, a
characteristic degeneracy appears in the $(\Omega_m,\Omega_\lambda)$ plane,
which is roughly orthogonal to the one given by the detection of high
redshift supernovae, and quite different from the CMB constraints.
A similar degeneracy was also found in the analog weak lensing analyses,
\citep{Lombardi} or by \citet{Gautret}.

\begin{figure}
  \resizebox{\hsize}{!}{\includegraphics{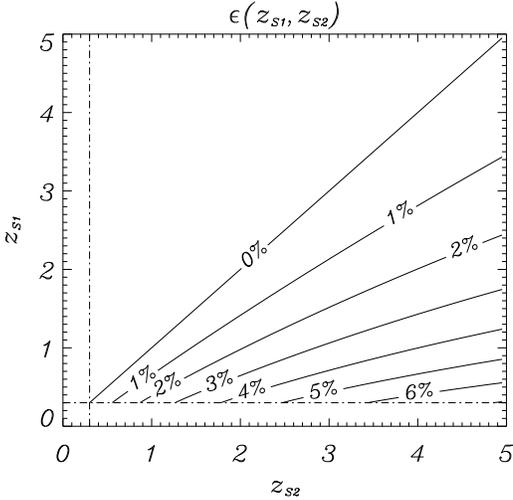}}
\caption{
  Relative difference between the ratio $e(z_{S1},z_{S2}) =
  E(z_{S1})/E(z_{S2})$ for two extreme cosmological models:
  $(\Omega_m=0.3, \Omega_\lambda=0)$ and ($\Omega_m=1,
  \Omega_\lambda=0$). This function $\varepsilon(z_{S1},z_{S2})$ is
  plotted in the $(z_{S1}, z_{S2})$ plane, assuming $z_{S1}\le z_{S2}$
  and $z_\mathrm{L}=0.3$.}
\label{Eps12}
\end{figure}

Another approach to quantify the dependence of a given lens configuration on
$\Omega_m$ and $\Omega_\lambda$ is to fix the lens redshift and to
search for two source redshifts $z_{S1}$ and $z_{S2}$ which give
the largest variation of the $E$-term when scanning the $(\Omega_m,
\Omega_\lambda)$ plane. For illustration we arbitrarily choose two
sets of cosmological parameters, for which the relative variation of $E$
is large: CP1 $(\Omega_m=0.3, \Omega_\lambda=0)$ and CP2 ($\Omega_m=1,
\Omega_\lambda=0$, {\it i.e.}  the EdS model).  Varying $z_{S1}$ and
$z_{S2}$, the function
$\varepsilon(z_{S1},z_{S2})=e_{CP2}(z_{S1},z_{S2})/e_{CP1}(z_{S1},z_{S2})-1$
represents the percentage of discrepancy between $CP1$ and $CP2$ for
$z_{S1}\geq z_{S2} \ (\geq z_\mathrm{L})$ (Fig.~\ref{Eps12}).

For a given high-redshift $z_{S2}$ source the best lowest source
redshift is $z_{S1} \simeq z_\mathrm{L}$, and for a given $z_{S1}$ the best
$z_{S2}$ is the highest redshift, the difference between cosmological models
increasing with $z_{S2}$. In all cases, this relative
difference is of the order of a few \%, meaning that the lens mass 
distribution must be known
to the same degree of accuracy to get further constraints
on the cosmological parameters. Hence, for 2 systems of images, the best
configuration is one background source close to the lens, in the
rising part of $E(z_\mathrm{S})$ and another one at high redshift, to
take into account the asymptotic value of the ratio. Note however that for
a source redshift close to the lens, the $E$-term becomes very small.
Also, the location of the images is very close to the lens center which makes
the detection of multiple images quite improbable, as small caustic
sizes imply small cross sections.

\subsection{Relative influence of the lens parameters}\label{lensparam}
\subsubsection{Physical assumptions}

In order to quantify the expected uncertainties on $\Omega_m$ and
$\Omega_\lambda$, it is possible to analytically estimate the
influence of the different lens models parameters. We use a
model of the potential derived from the mass density described by
\citet{Hjorth}, hereafter HK. It is based on a physical scenario of
violent relaxation in galaxies, also valid in clusters of galaxies.
The mass density is characterized by a core radius $a$ and a cut-off
radius $s$:
\begin{equation}
\rho(r)=\frac{\rho_{0}}{\left (1+\frac{r^{2}}{a^{2}}\right )
\left (1+\frac{r^{2}}{s^{2}}\right )}.
\label{hjorthkneib}
\end{equation}
Then the projected mass density $\Sigma(\theta)$ is represented by: 
\begin{equation}
\Sigma(\theta)=\Sigma_0\ \frac{\theta_a\theta_s}{\theta_s-\theta_a}
\ \left(\frac{1}{\sqrt{\theta^2+\theta_a^2}}-
\frac{1}{\sqrt{\theta^2+\theta_s^2}}\right),
\end{equation}
where $\theta$ is the angular coordinate, $\theta_a = a/D_\mathrm{OL}$ and
$\theta_s = s/D_\mathrm{OL}$. $\Sigma_0$ is a normalization factor,
related to the cluster parameters: 
\begin{equation}
\Sigma_0 = \pi \rho_0 D_\mathrm{OL} \ \frac{\theta_a\theta_s}{\theta_s+\theta_a}.
\end{equation}
Finally,the mass inside the projected angular radius $\theta$ is:
\begin{eqnarray}
M(\theta) & = & 2\pi\Sigma_0 D_\mathrm{OL}^2 \ 
\frac{\theta_a\theta_s}{\theta_s-\theta_a} \times \nonumber \\ 
& & \quad \left[\sqrt{\theta^2+\theta_a^2}-\sqrt{\theta^2+\theta_s^2}+
\theta_s-\theta_a\right]. 
\label{dens}
\end{eqnarray}

The velocity dispersion $\sigma(\theta)$ is related to the mass
density and the gravitational potential via the Jeans equation.
Assuming an isotropic velocity dispersion and retaining terms up to
first order in $\theta_a/\theta_s$, we get the relation between the
central velocity dispersion $\sigma_0=\sigma(0)$ and $\rho_0$:
\begin{equation}
\sigma_0^2 = \frac{\pi^3Ga^2}{2} \ \rho_0 s.
\end{equation}
Finally we compute the expression of the deviation angle between the
positions of the source and of the image due to the lens:
$D_{\theta_{I}}=\|\vec{\theta_{I}}-\vec{\theta_{S}}\|$, neglecting
second order terms in $\theta_a/\theta_s$ (we suppose here $s\gg a$):
\begin{equation}
D_{\theta_{I}} = \frac{16}{\pi}\frac{\sigma_0^2}{c^2}\,E\,
\frac{1}{\theta_I} \left[\sqrt{\theta^2+\theta_a^2}
-\sqrt{\theta^2+\theta_s^2}+ \theta_s-\theta_a\right].
\label{dev}
\end{equation}

\subsubsection{The single multiple-image configuration}
\begin{figure}
  \resizebox{\hsize}{!}{\includegraphics{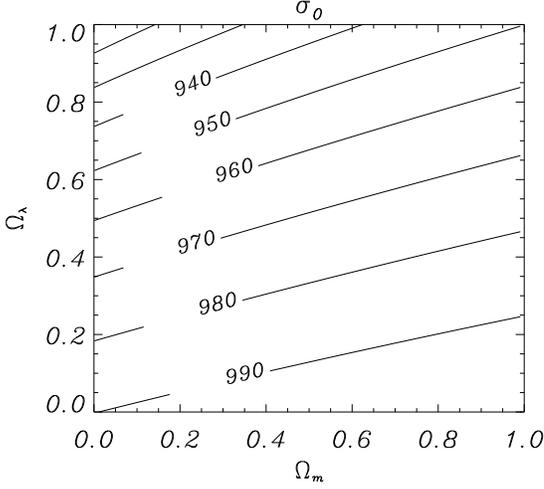}}
\caption{
  Variation of the central velocity dispersion $\sigma_0$ in the
  ($\Omega_m,\Omega_\lambda$) plane, assuming that the product
  $\sigma_0^2 E(z_\mathrm{S})$ is constant. $\sigma_0$ has been fixed to 1000
  km~s$^{-1}$ for $\Omega_m=1$, $\Omega_\lambda=0$ while $z_\mathrm{L}=0.3$ and
  $z_\mathrm{S}=1$. }
\label{Vdisp}
\end{figure}

The central velocity dispersion (or equivalently the mass
normalization of the cluster core) is obviously the predominant factor
in any lens configuration. With a single family of images we can only
constrain the combination $\sigma_0^2E$ and cannot disentangle the
influence of the cosmological parameters and the absolute normalization
of the mass (Fig.~\ref{Vdisp}). If we were able to measure the total
mass within the Einstein radius independently from lensing techniques
and with an accuracy better than a few \%, we could in principle put
some constraints on $\Omega_\lambda$. Observationally, there are 2
situations where it is likely that we could disentangle the
effect of cosmology and absolute mass:\\
1) in the case of a cluster-lens with extremely good X-ray
data, particularly in estimating the temperature distribution of the gas
(under the assumption of hydrostatic equilibrium),\\
2) in the case of a multiple system around a single galaxy, for which
one is able to measure accurately the stellar velocity dispersion of
the lensing galaxy \citep{Tonry}. \\ 
However in both cases this
represents some observational challenge and requires the most
powerful instruments to achieve this goal.

Although the error budget in the image positions is dominated by the
error on the total cluster mass (or equi\-valently the velocity
dispersion), we can determine the relative influence of the other
parameters to infer the importance of $\Omega_m$ and $\Omega_\lambda$
in the image formation. The relative error on the deviation angle
$D_{\theta_I}$ depends on $\sigma_0$, $\theta_a$, $\theta_s$ and
$\theta_I$ for the gravitational potential, and $z_\mathrm{L}$, $z_\mathrm{S}$,
$\Omega_m$ and $\Omega_\lambda$ for the $E$-term (Eq.(\ref{dev})). 
Therefore we can
write:

\begin{equation}
\frac{\mathrm{d}D_{\theta_I}}{D_{\theta_I}}=\frac{\mathrm{d}E}{E}+
\alpha_{\sigma_0}\frac{\mathrm{d}\sigma_0}{\sigma_0}+
\alpha_{\theta_a}\frac{\mathrm{d}\theta_a}{\theta_a}+
\alpha_{\theta_s}\frac{\mathrm{d}\theta_s}{\theta_s}+
\alpha_{\theta_I}\frac{\mathrm{d}\theta_I}{\theta_I},
\end{equation}
with
\begin{equation}
\frac{\mathrm{d}E}{E} (z_\mathrm{S}) = \alpha_{\Omega_m}\frac{\mathrm{d}\Omega_m}
{\Omega_m}+
\alpha_{\Omega_\lambda}\frac{\mathrm{d}\Omega_\lambda}{\Omega_\lambda}+
\alpha_{z_\mathrm{L}}\frac{\mathrm{d}z_\mathrm{L}}{z_\mathrm{L}}+\alpha_{z_\mathrm{S}}\frac{\mathrm{d}z_\mathrm{S}}{z_\mathrm{S}}.
\end{equation}

$\alpha_{\theta_a}$, $\alpha_{\theta_s}$ and $\alpha_{\theta_I}$ can
be computed analytically while $\alpha_{\sigma_0}=2$ is the largest
factor.  Since the angular diameter distances do not have an analytic
expression if $\Omega_\lambda$ is non-zero, the coefficients
$\alpha_{\Omega_m}$, $\alpha_{\Omega_\lambda}$, $\alpha_{z_\mathrm{L}}$ and
$\alpha_{z_\mathrm{S}}$ must be computed numerically. In practise, they are
computed for a given set of parameters $(z_\mathrm{L}, \Omega_m,
\Omega_\lambda)$ as their variation with $\Omega_m$ and
$\Omega_\lambda$ is of higher order.  For a reasonable set of lens
parameters, the $\alpha$-coefficients are of the same order of
magnitude, except that $\alpha_{z_\mathrm{S}}$ and $\alpha_{z_\mathrm{L}}$ can dominate
the error budget if the source redshift is close to the lens (an unlikely
case). On the contrary, $\alpha_{\Omega_m}$ and $\alpha_{\Omega_\lambda}$ are 
of second order, and $\alpha_{\Omega_\lambda}$ is somewhat larger than
$\alpha_{\Omega_m}$. This reflects again the fact that $E$-term is
more sensitive to $\Omega_\lambda$ than to $\Omega_m$.

To quantify the relative influence of all the parameters in the case
of a single family of images, we computed explicitely
$\mathrm{d}D_{\theta_I}/D_{\theta_I}$ in two cases, for a cluster-lens and for
a galaxy-lens.

\noindent 1) For a cluster of galaxies, we take the following parameters: 
$z_\mathrm{L}=0.3$,
$z_\mathrm{S}=4$, $\Omega_m=0.3$, $\Omega_\lambda=0.7$, $\theta_s/\theta_a=10$
and $\theta_I/\theta_a=4$. We thus find from Eq.(\ref{dev}):
\begin{eqnarray}
\frac{\mathrm{d}D_{\theta_I}}{D_{\theta_I}} & = & 2 \, \frac{\mathrm{d}
\sigma_0}{\sigma_0} 
- 0.31 \, \frac{\mathrm{d}\theta_I}{\theta_I} - 0.21 \,
\frac{\mathrm{d}\theta_a}{\theta_a}
+ 0.51 \, \frac{\mathrm{d}\theta_s}{\theta_s} \nonumber \\
 & - & 0.17 \, \frac{\mathrm{d}z_\mathrm{L}}{z_\mathrm{L}} + 0.062 \, \frac{\mathrm{d}z_\mathrm{S}}{z_\mathrm{S}} 
     + 0.012 \, \frac{\mathrm{d}\Omega_{m}}{\Omega_{m}} + 0.14 \,
     \frac{\mathrm{d}\Omega_{\lambda}}{\Omega_{\lambda} \nonumber}\\
 & &  
\label{erreur_D}
\end{eqnarray}

Let us assume a perfectly known mass profile (i.e. d$\theta_a=\mathrm{d}
\theta_s=0.$). Neglecting the influence of $\Omega_m$, we ask what precision
would be required on $\sigma_0$ to derive an error of 50\% on 
$\Omega_{\lambda}$. 
The accuracy of the position of the center of the images is calculated using 
the first moment of the flux $f(\theta)$ on a
given image: $\theta_I=\int \theta f(\theta)\mathrm{d}\theta^2 /\int
f(\theta)\mathrm{d}\theta^2$ which yields an error $\mathrm{d}
{\theta_I}$ of a fraction
of the spatial resolution. HST observations are then required to reach
$\mathrm{d}\theta_I=0.1\arcsec$ (LP98) or better. To reduce the 
uncertainty on the 
redshift measurements, we assume spectroscopic determinations, so that
$dz \simeq0.001$. Finally we have to compute
the relative errors on $D_{\theta_I}=
\Vert \vec \theta_I - \vec \theta_S \Vert$, so the position of the
source is in principle required. But as we are in the strong lensing
regime, we assume that $\theta_S\ll\theta_I$, so that both
quantities $D_{\theta_I} \simeq \theta_I$ and
$\displaystyle{\frac{\mathrm{d}D_{\theta_I}}
  {D_{\theta_I}} \simeq 
\frac{\mathrm{d}\theta_I}{\theta_I}}$ are directly related to
observable ones. Taking these values into account, we need to know $\sigma_0$
with 3.6\% accuracy to get the expected constraint on $\Omega_{\lambda}$.
Such an accuracy is out of reach with observations of clusters of 
galaxies.

\noindent 2) For a single galaxy, we consider typically: $z_\mathrm{L}=0.3$,
$z_\mathrm{S}=4$, $\Omega_m=0.3$, $\Omega_\lambda=0.7$, $\theta_s/\theta_a=200$
and $\theta_I/\theta_a=200$ (ratios taken from the modelisation of the lens
HST 14176+5226 by \citet{Hjorth}), leading to:
\begin{eqnarray}
\frac{\mathrm{d}D_{\theta_I}}{D_{\theta_I}} & = & 2 \, \frac{\mathrm{d}
\sigma_0}{\sigma_0} 
- 0.72 \, \frac{\mathrm{d}\theta_I}{\theta_I} - 0.0066 \,
\frac{\mathrm{d}\theta_a}{\theta_a}
+ 0.73 \, \frac{\mathrm{d}\theta_s}{\theta_s} \nonumber \\
 & - & 0.17 \, \frac{\mathrm{d}z_\mathrm{L}}{z_\mathrm{L}} + 0.062 \, \frac{\mathrm{d}z_\mathrm{S}}{z_\mathrm{S}} 
     + 0.012 \, \frac{\mathrm{d}\Omega_{m}}{\Omega_{m}} + 0.14 \,
     \frac{\mathrm{d}\Omega_{\lambda}}{\Omega_{\lambda} \nonumber} \\
 & &
\label{erreur_D2}
\end{eqnarray}

Taking the same values for the observational errors and considering a 
perfectly known mass 
profile, we require an accuracy of 6.4\% on $\sigma_0$ to 
derive a 50\% error on $\Omega_{\lambda}$. For a typical galaxy, this 
represents 
about 15 km~s$^{-1}$. \citet{Warren} measured the 
velocity dispersion in the deflector of the Einstein ring 
0047--2808 with an error of 30  km~s$^{-1}$. 
A better accuracy could be obtained
by looking at particular strong absorption features with 10m class telescope
observations. This could be sufficient to confirm an accelerating 
Universe.

\subsubsection{Configuration with 2 multiple-image systems}
With a second system of multiple images another region of the $E(z_\mathrm{S})$
curve is probed while the cluster parameters are the same. In that case,
the relevant quantity becomes the ratio of the deviation angles for 2
images $\theta_{I^1}$ and $\theta_{I^2}$ belonging to 2 different
families at redshifts $z_{S1}$ and $z_{S2}$. We define
$R_{\theta_{I^1},\theta_{I^2}}=\displaystyle{\frac{D_{\theta_{I^2}}}
  {D_{\theta_{I^1}}}}$. This function has the advantage that it does not
depend on $\sigma_0$ anymore.  Following our previous definitions,
we can write
\begin{eqnarray}
\label{error2}
\frac{\mathrm{d}R_{\theta_{I^1},\theta_{I^2}}}{R_{\theta_{I^1},\theta_{I^2}}} 
 & = & \frac{\mathrm{d}E(z_{S2})}{E(z_{S2})} - \frac{\mathrm{d}E(z_{S1})}
{E(z_{S1})}
 \nonumber \\
 & + & \alpha_{\theta_I}\left(\theta_{I^2}\right) \, 
\frac{\mathrm{d}\theta_{I^2}}{\theta_{I^2}}  
- \alpha_{\theta_I}\left(\theta_{I^1}\right) \,
\frac{\mathrm{d}\theta_{I^1}}{\theta_{I^1}} \nonumber \\
 & + & \Big( \alpha_{\theta_a}\left(\theta_{I^2}\right)
- \alpha_{\theta_a} \left(\theta_{I^1}\right) \Big) \, 
\frac{\mathrm{d}\theta_a}{\theta_a} \nonumber \\
 & + & \Big( \alpha_{\theta_s}\left(\theta_{I^2} \right) 
- \alpha_{\theta_s} \left(\theta_{I^1}\right) \Big) \, 
\frac{\mathrm{d}\theta_s}{\theta_s}
\end{eqnarray}

Numerically, we chose a typical configuration to compute
$\displaystyle{\frac{\mathrm{d}R_{\theta_{I^1},\theta_{I^2}}}
  {R_{\theta_{I^1},\theta_{I^2}}}}$: $\theta_{s} / \theta_{a} =10$,
$\theta_{I^2} / \theta_{a}=4$, $\theta_{I^2} / \theta_{I^1}=2$,
$z_\mathrm{L}=0.3$, $z_{S1}=0.6$, $z_{S2}=5$, assuming $\Omega_m=0.3$ and
$\Omega_\lambda=0.7$. This gives the following error budget:
\begin{eqnarray}
\frac{\mathrm{d}R_{\theta_{I^1},\theta_{I^2}}}{R_{\theta_{I^1},\theta_{I^2}}} 
& = & 0.92 \, \frac{\mathrm{d}z_{L}}{z_{L}} - 0.99 \, \frac{\mathrm{d}z_{S1}}
{z_{S1}} 
+ 0.062 \, \frac{\mathrm{d}z_{S2}}{z_{S2}} \nonumber \\
& - & 0.018 \, \frac{\mathrm{d}\theta_{I^1}}{\theta_{I^1}} 
- 0.31 \, \frac{\mathrm{d}\theta_{I^2}}{\theta_{I^2}} 
+ 0.12 \, \frac{\mathrm{d}\theta_{a}}{\theta_{a}} \nonumber \\
& + & 0.21 \, \frac{\mathrm{d}\theta_{s}}{\theta_{s}} 
+ 0.034 \, \frac{\mathrm{d}\Omega_{m}}{\Omega_{m}} 
+ 0.037 \, \frac{\mathrm{d}\Omega_{\lambda}}{\Omega_{\lambda}} 
\end{eqnarray}

The contribution of the physical lens parameters in this error budget
is strongly attenuated comprared to the single family case. There is no 
more dependence on $\sigma_0$ and
the dependence on the mass profile ( $\theta_a, \theta_s$) is reduced
by about a factor of 2 compared to a single family of images. This
corresponds to the variation of the potential between $\theta_{I1}$
and $\theta_{I2}$, the absolute normalization being removed. Anyhow,
this can still represent the main source of error because we cannot
expect to constrain $\theta_a$ to better than 1.5\% and $\theta_s$ to
better than 2\%\ typically (see Sect.~\ref{number_syst}).

For the source redshifts, we have selected one of the sources at
$z_{S1} = 0.6$, which means that its
$\alpha$-coefficient is quite large. The accurate value of the
redshifts is thus fundamental, and a spectroscopic determination is
essential ($dz \simeq0.001$). A photometric redshift estimate would
not be satisfactory, because we cannot expect an accuracy better than
10\%\ in most cases ($\mathrm{d}z \simeq 0.1 - 0.2$, \citet{Bolzonella}). We 
keep
$\mathrm{d}\theta_I=0.1\arcsec$. The strong lensing regime approximation 
leads to
$R_{\theta_{I^1},\theta_{I^2}}=\displaystyle{
  \frac{\Vert \vec \theta_{I^2} - \vec \theta_{S^2} \Vert} {\Vert \vec
    \theta_{I^1} - \vec \theta_{S^1} \Vert}}\simeq
\displaystyle{\frac{\theta_{I^2}}{\theta_{I^1}}}$ and
$\displaystyle{\frac{\mathrm{d}R_{\theta_{I^1},\theta_{I^2}}}
  {R_{\theta_{I^1},\theta_{I^2}}} \simeq 
\frac{\mathrm{d}\theta_{I^2}}{\theta_{I^2}}
  - \frac{\mathrm{d}\theta_{I^1}}{\theta_{I^1}}}$.

\begin{figure*}
  \hspace{1cm}
  \resizebox{\hsize}{!}{
\includegraphics{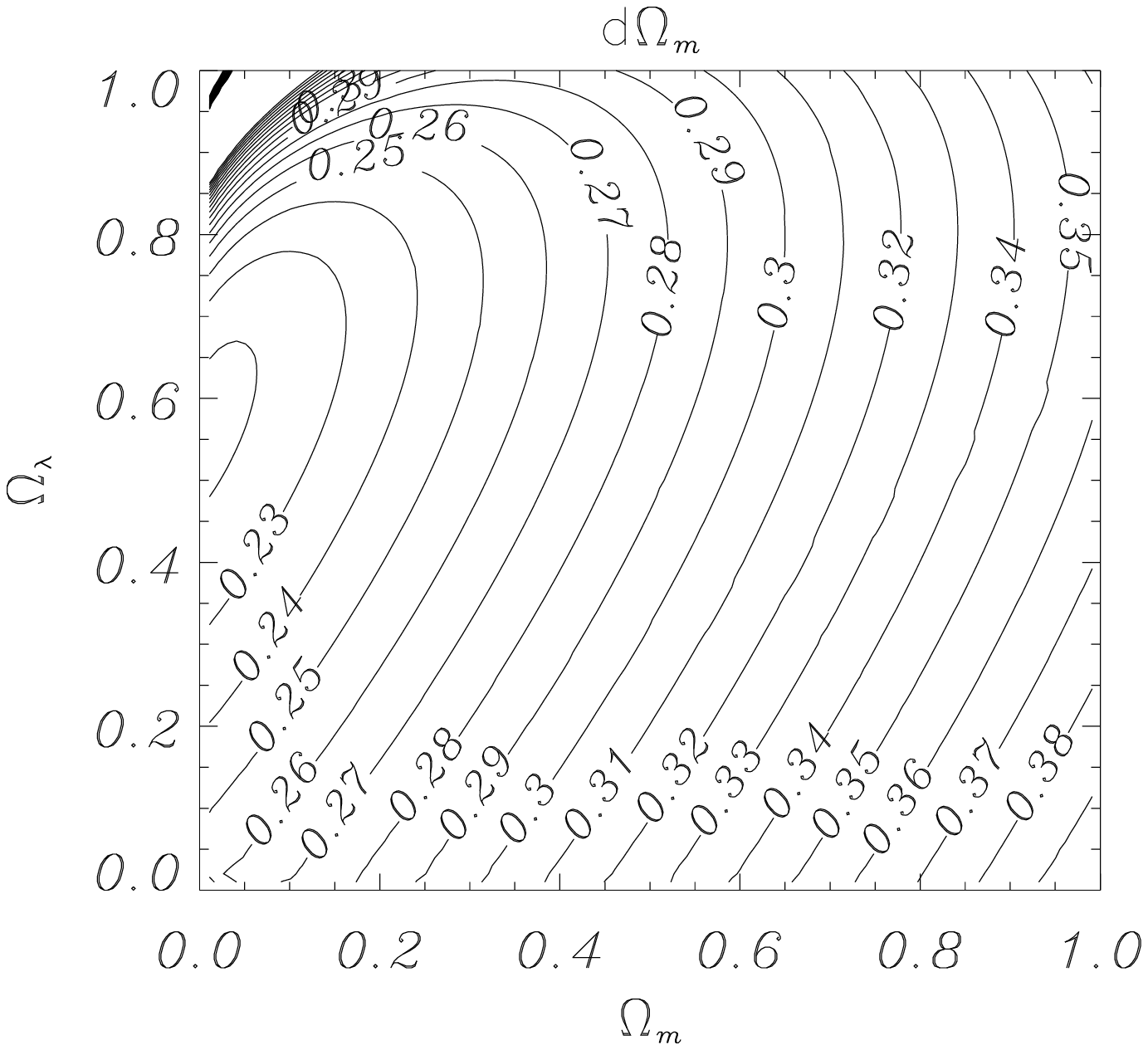}
\includegraphics{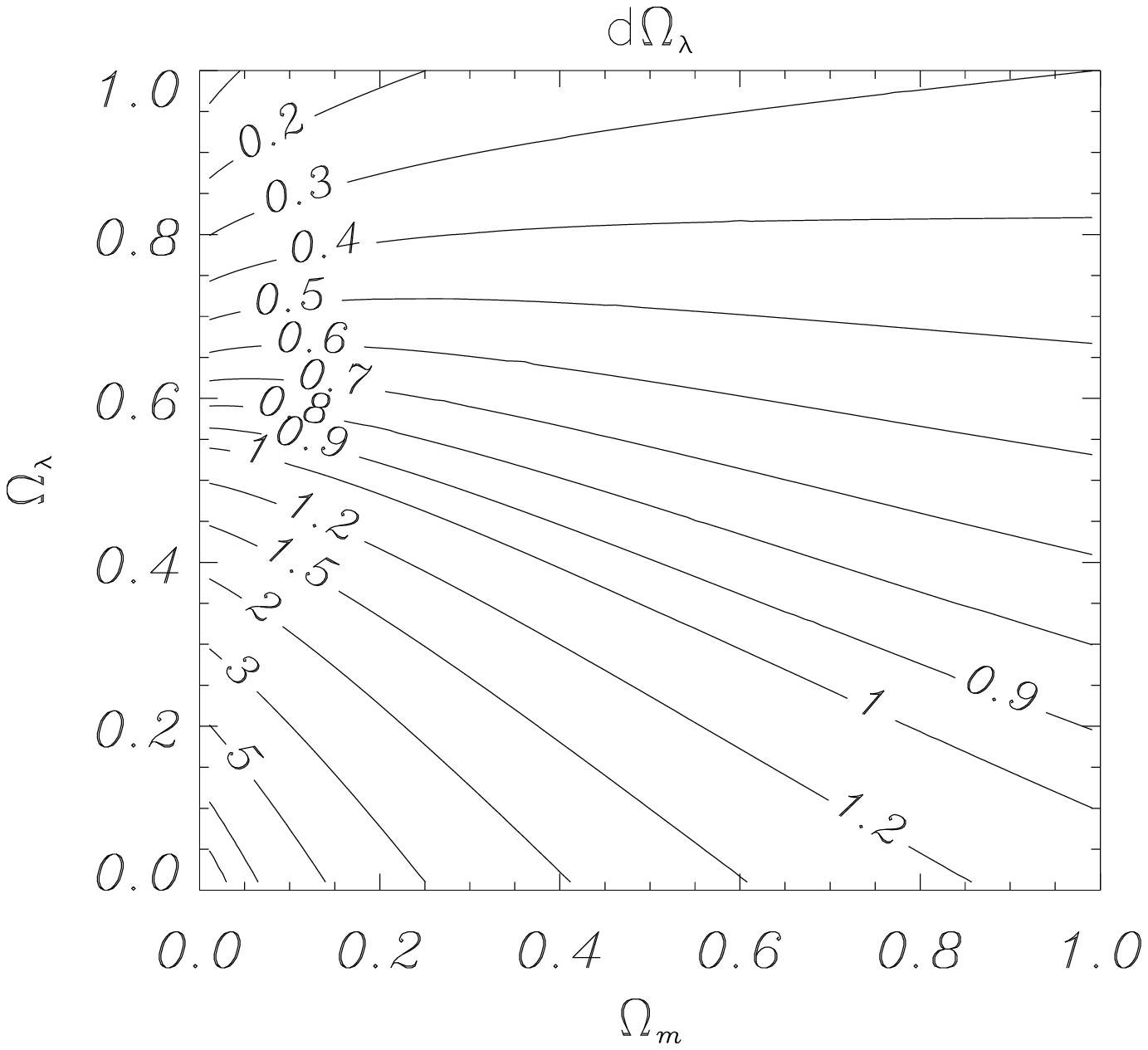}}
\caption{
  Final errors $d\Omega_m$ {\bf (Left)} and $d\Omega_\lambda$ {\bf
    (Right)} for a given $(\Omega_m, \Omega_\lambda)$ in the lens
  configuration discussed in the text (Section \ref{lensparam}), for
  two source redshifts $z_{S1}=0.6$ and $z_{S2}=5$. }
\label{Err_omla}
\end{figure*}

We can then separate the contributions of the parameters that do not
depend on $\Omega_m$ or $\Omega_\lambda$ from those which depend on
them and re-write Eq.(\ref{error2}):
\begin{eqnarray}
& & A_{\Omega_m} \frac{\mathrm{d}\Omega_m}{\Omega_m} + 
A_{\Omega_\lambda} \frac{\mathrm{d}\Omega_\lambda}{\Omega_\lambda} = \\
& & \sqrt{\mathrm{Err1}^2 \left( \theta_{I^1},\theta_{I^2}, \theta_a,
    \theta_s \right) + \mathrm{Err2}^2 \left( \Omega_m,
\Omega_\lambda, z_\mathrm{L}, z_{S1}, z_{S2} \right)}\nonumber
\end{eqnarray}
$A_{\Omega_m}$ and $A_{\Omega_\lambda}$ depend on $\Omega_m,
\Omega_\lambda, z_\mathrm{L}, z_{S1}, z_{S2}$ while Err1$^2$ and Err2$^2$ are
the quadratic sums of the errors, with a separation between the
geometrical parameters and those depending on the cosmology.  For each
set of cosmological parameters we then compute all these coefficients
numerically. In addition, we also need a calculation of the
``degeneracy'' $\partial \Omega_m/ \partial\Omega_{\lambda}$ to obtain
either $\mathrm{d}\Omega_m$ or $\mathrm{d}\Omega_{\lambda}$. This is the 
slope of the
degeneracy curves of the E-terms ratio plotted in Fig.~\ref{E2omla}.
Indeed considering 2 points $(\Omega_m,\Omega_\lambda)$ and
$(\Omega_m+\mathrm{d}\Omega_m,\Omega_\lambda+ \mathrm{d}\Omega_\lambda)$ on 
such a curve
(for a given set of $z_\mathrm{L}$, $z_{S1}$, $z_{S2}$), we have
$e(\Omega_m,\Omega_\lambda)=e(\Omega_m+\mathrm{d}\Omega_m,\Omega_\lambda+
\mathrm{d}\Omega_\lambda)$ so that we get $\partial \Omega_m/ \partial
\Omega_{\lambda}=-\partial_{\Omega_{\lambda}}
e(z_{S1},z_{S2})/\partial_{\Omega_m}e(z_{S1},z_{S2})$. The final
expected errors on $\mathrm{d}\Omega_m$ and $\mathrm{d}\Omega_\lambda$ are 
plotted in
Fig.~\ref{Err_omla} for a continuous set of world models. The
method is in general far more sensitive to the matter density than to
the cosmological constant, for which the error bars are larger. This
apparent contradiction with the general statement that lensing is more
sensitive to the cosmological constant than to the matter density is
due to the fact that we analysed the {\bf ratio} of two E-terms and
this ratio varies more rapidly with $\Omega_m$ when scanning the
$(\Omega_m, \Omega_\lambda)$ plane (Fig~\ref{E2omla}). For
illustration, we quantitatively obtain the following errors for the
corresponding cosmological models:
\begin{eqnarray*}
\Lambda\mathrm{CDM}: & \qquad & \delta \Omega_m=0.11  
\qquad \delta \Omega_\lambda=0.23 \\
\mathrm{SCDM}: & \qquad & \delta \Omega_m=0.17 
\qquad \delta \Omega_\lambda=0.48
\end{eqnarray*}

This analysis shows that the expected results are quite encouraging,
and the constraints we could get are similar to the ones currently
obtained by other methods. Note however that these typical values
require both HST imaging of cluster lenses and deep spectroscopic
observations for the redshift determination of multiple arcs. They may
depend on the choice of the lens parameters and on the potential
model chosen to describe the lens, a problem that we will now investigate.


\section{Constraints on the cosmological parameters from strong
  lensing}

\subsection{Existence of multiple systems of lensed images}
\label{mult_sys}
More and more cluster-lenses are known to show several systems of
multiple images (with spectroscopic or photometric redshifts). Lens
modelling is then performed with a good accuracy and allows the
prediction of extra families of images and their expected redshifts.
When these images are later identified and if their redshift can be
measured spectroscopically, an iterative process brings the lens model
to a high level of accuracy, where most of the parameters which
characterise the mass profile are strongly constrained. This full
process has been applied successfully in a few clusters such as A2218 
\citep{Kneib3}, A370
\citep{Kneib2, Smail, Bezecourt2} or AC114 \citep{Natarajan98,Campusano}.

In order to apply more systematically the method proposed here, we may
ask whether the few cited cluster-lenses are representative of some
generic cluster or if they correspond to very peculiar configurations.
To answer this question, we simulated a typical cluster at redshift
$z=0.2$ with the following characteristics. A main clump is described
with the potential of Eq.(\ref{hjorthkneib}), the so-called HK mass
density, with $a=50$~kpc and $s=500$~kpc. These values are typical of
cluster-lenses at this redshift \citep{Smith}. The central velocity
dispersion is varied from 800 to 1400 km s$^{-1}$ to allow a variation
of the Einstein radius. In addition, 12 individual galaxies are added
in the mass distribution, following the prescription used by
\citet{Natarajan} and for a total contribution of 30\% of the total
mass. Their individual masses are scaled with respect to their
luminosity $L_i$ by the Faber-Jackson relation:
\begin{equation} 
\sigma_i=\sigma_0^G \left(\frac{L_i}{L_0}\right)^{\frac{1}{4}} 
\end{equation}
with $\sigma_0^G=150$ km s$^{-1}$ (following
\citet{Faber}), and with a cut-off radius:
\begin{equation}
\theta_{S_i}=\theta_{S_0}^G \left(\frac{L_i}{L_0}\right)^{\frac{1}{2}} 
\end{equation}
providing a constant ratio $M/L$ \citep{Natarajan97}.

To simulate the background sources, we used the Hubble Deep Field
(HDF) image acquired by the {\textit{HST}} \citep{Williams}.  From
\citet{Fernandez}, 946 galaxies were extracted from the deepest zone
of the F814W image, up to a magnitude limit $AB(8140)=28.0$ and over
an angular area of 3.92 arcmin$^2$. These authors also provide a
catalog of photometric redshifts for all these objects. In addition,
for about 10\% of them, a spectroscopic redshift is available.  We
used this redshift distribution (spectroscopic redshift preferably
used when available) as a sample of galaxy-sources to be lensed by the
simulated clusters. In order to increase the statistical significance of this
simulation, we generated a source catalogue with 10 times the number 
of galaxies extracted from the HDF image.
We then distributed these sources at random angular positions over
the central inner $40\times40$\ arcsec$^2$. We checked that this
region includes the external radial caustic line, so that no multiple
images are lost. The increase in the galaxy density
is then corrected for in the final results.

\begin{table}
\caption[]{Number of systems of images obtained for simulated cluster 
potentials with different 
values of the central velocity dispersion $\sigma_0$ (the 
corresponding Einstein radius $R_E$ is given for $z=1$). The redshift
distribution of the sources is assumed 
from the HDF data. We take random 
positions for the sources over the central inner $40\times40$~arcsec$^2$.
$n_j$ is the number of systems of $j$ images. $n_j^*$ is the number of
systems of $j$ images with $AB(8140)<24.5$, corresponding to
``observable'' ones. Then, each system is counted both in $n_i$ and in
$n_j^*$ with $j\le i$ (if only $j$ images among $i$ are detectable). Systems
counted in $n_0$ show no ``observable'' image. So $n_1+n_\mathrm{tot}=n_0+
n_1^*+n_\mathrm{tot}^*$, which is the number of galaxies in the selected 
field.}
\label{table_mul}
\begin{flushleft}
\begin{tabular}{cclrrrr}
\hline\noalign{\smallskip}
$\sigma_0$   & \multicolumn{2}{c}{(km s$^{-1}$)} & 800   & 1000 & 1200 & 1400\\
\noalign{\smallskip}
\hline\noalign{\smallskip}
\noalign{\smallskip}
$R_E$        & \multicolumn{2}{c}{(arcsec)}         &  5    &   14 &  28  &   40\\
\noalign{\smallskip}
\hline\noalign{\smallskip}
\noalign{\smallskip}
             & 0                       &   $n_0^{\ \,(1)}$ &  78  &   73 &   69 &   65\\
                                     & & $n_0^*$ &  0   &   0 &   0 &   0\\
\noalign{\smallskip}
\cline{2-7}  \noalign{\smallskip} & 1  & $n_1^{\ \,(2)}$ & 107   &  107 &   99 &   66\\
                                     & & $n_1^{*\,(3)}$ &  29   &   34 &   34 &   26\\
\noalign{\smallskip}
\cline{2-7}  \noalign{\smallskip} & 2  &   $n_2$ &   0   & 0.068& 0.14 & 0.10\\
                                     & & $n_2^*$ & 0.11  &  0.41&  2.3 &   13\\
\noalign{\smallskip}
\cline{2-7}  \noalign{\smallskip} & 3  &   $n_3$ & 0.12  &  0.60&  8.0 &   41\\
                                     & & $n_3^*$ &0.034  & 0.068&  1.7 &  3.9\\
\noalign{\smallskip}
\cline{2-7} \noalign{\smallskip} $j$ & 4 &   $n_4$ &0.034  & 0.011&  0.034&0.057\\
                                     & & $n_4^*$ &0.011  & 0.011&  0.034&0.011\\
\noalign{\smallskip}
\cline{2-7}  \noalign{\smallskip} & 5  &   $n_5$ &0.022  & 0.011& 0.11&0.011\\
                                     & & $n_5^*$ &0.011  & 0    &  0   &0.011\\
\noalign{\smallskip}
\cline{2-7}  \noalign{\smallskip} & 6  &   $n_6$ &0.011  & 0    & 0    &0.011\\
                                     & & $n_6^*$ &0      & 0    & 0    &0.011\\
\noalign{\smallskip}
\cline{2-7}  \noalign{\smallskip} & 7  &   $n_7$ &0.011  & 0    & 0.011&0.011\\
                                     & & $n_7^*$ &0      & 0.011& 0    &0    \\
\noalign{\smallskip}
\cline{2-7}  \noalign{\smallskip} & 8  &   $n_8$ &0      & 0.011& 0    &0    \\
                                     & & $n_8^*$ &0      & 0    & 0    &0    \\
\noalign{\smallskip}
\hline\noalign{\smallskip}
total        &  $(j>2)$    &$n_\mathrm{tot}$   & 0.18  & 0.70 &  8.3 &   41\\
             &             &$n_\mathrm{tot}^*$ & 0.17  & 0.50 &  4.0 &   17\\
\noalign{\smallskip}
\hline
\end{tabular}
\begin{tabular}{ll}
$^{(1)}$including 0.6 galaxies at $z\le 0.2$.\\
$^{(2)}$including 5.2 galaxies at $z\le 0.2$.\\
$^{(3)}$including 4.6 galaxies at $z\le 0.2$.\\
\end{tabular}
\end{flushleft}
\end{table}

Table \ref{table_mul} presents for each value of the central velocity
dispersion the number of systems found with their image multiplicity.
We also determined the number of systems in which each image could be
observed (with a magnitude $AB(8140)<24.5$, corresponding to typical
HST integration time of 10 ksec). Objects with
$AB(8140)>28.0$ could be observed due to the lens effect
if the magnification exceeds a factor of 25. This
very rare configuration is neglected in our simulations for simplicity. For a
cluster massive enough ($\sigma_0\ge 1200$ km s$^{-1}$, corresponding
to $M_\mathrm{tot}\ge 2.10^{14}M_{\sun}$ for our potential model), numerous systems
of multiple images (mainly triple images) are formed and a significant
fraction could be observable. Although these simulations are quite
simple and cannot be used for realistic statistics of image formation,
it gives us confidence that the use of multiple image families for the
determination of the cosmological parameters is achievable and
should be applied on a large number of rich clusters.

\subsection{Method and algorithm for numerical simulations}
In most cases, clusters of galaxies present a global ellipticity in
their light distribution or in their gas distribution traced by X-ray
isophotes. It is generally believed that this is related to an
ellipticity in the mass distribution. This has indeed been recognized
several times by the modeling of cluster lenses such as MS2137--23
\citep{Mellier2} or Abell 2218 \citep{Kneib3}. So we include such an
ellipticity in our modeling of cluster potentials. The basic
distribution of matter we consider is again the HK one, with, in
addition, a substitution of the radial distance $r$ by $R$ defined as:
\begin{equation}
R =
\left(\frac{X\cos\theta+Y\sin\theta}{1+\epsilon}\right)^2 + 
\left(\frac{-X\sin\theta+Y\cos\theta}{1-\epsilon}\right)^2
\label{R}
\end{equation}
where $X=(x-x_0)$ and $Y=(y-y_0)$. The potential $\phi$ is then
characterized by 7 parameters, namely: $x_0, y_0, \epsilon, \theta$
for the geometry of the lens and $\sigma_0, \theta_a, \theta_s$ for
the shape of the mass profile.

Another popular density profile to be tested is the so-called Navarro,
Frenk \& White (NFW) density profile found in many simulations of dark
matter and cluster formation \citep{NFW}:

\begin{equation}
\rho(r)=\frac{\rho_{\mathrm c}}{(r/r_s)
(1+r/r_s)^2}
\label{rho_nfw}
\end{equation}
\noindent where $\rho_{\mathrm c}$ is a characteristic density and $r_s$ a scale
radius. No analytic developments
have been proposed so far for the corresponding ellipsoidal profile.
In a companion paper \citep{GolseNFW} we propose a new 
``pseudo-elliptical'' NFW profile and
compute its lensing properties. The corresponding
potential is characterized by 6 parameters: $x_0, y_0, \epsilon,
\theta$ for the geometry of the lens and $v_c, \theta_s$ for the
shape of the mass profile. The characteristic velocity $v_c$ is defined by

\begin{equation}
v_c^2=\frac{8}{3}\mathrm{G}r_s^2\rho_c
\end{equation}

\noindent as explained in \citet{GolseNFW}.

To create a simulated lens configuration we need to fix some arbitrary
values of the cosmological parameters
$(\Omega_m^0$,$\Omega_\lambda^0)$ as well as the cluster lens redshift
$z_\mathrm{L}$. The numerical code {\it LENSTOOL} developed by one of us
\citep{Kneib1} can then trace back the source of a given image or
determine the images of an elliptical source galaxy at a redshift
$z_\mathrm{S}$. The initial data are several sets of multiple images at
different redshifts. In all cases we do not take into account the
central de-magnified images, which are generally not detected. With
these observables, we can recover some parameters of the potential
while we scan a grid in the $(\Omega_m$,$\Omega_\lambda)$ plane.  The
likelihood of the result is obtained via a $\chi^2$-minimization (with
a parabolic or a Monte Carlo method), where $\chi^2$ is computed in
the source plane as:  

\begin{equation}
\chi^2=\displaystyle{\sum_{i=1}^n \sum_{j=1}^{n^i}
\frac{[\mathcal A_j^i(\vec{\theta_{S_j^i}}-\vec{\theta_{SG^i}})]^2}
{\sigma_{I_j^i}^2}}
\label{chi2}
\end{equation}
The superscript $i$ refers to a given family of multiple images and
the subscript $j$ to the images inside a family of $n_i$ images.
There is a total of $\sum_{i=1}^n n_i=N$ images, and $\sum_{i=1}^n
2(n_i-1)=N_C$ constraints on the models assuming that only the
position of the images are fitted. $\vec{\theta_{S_j^i}}$ is the
source position associated with the image $\vec{\theta_{I_j^i}}$ in the lens
inversion. $\vec{\theta_{SG^i}}$ is the barycenter of all the
$\vec{\theta_{S_j^i}}$ belonging to the same family $i$. $\mathcal A_j^i$ is
the magnification matrix for a particular image 
and $\sigma_{I_j^i}$ is the error on the
position of the center of image $\vec{\theta_{I_j^i}}$.
Quantitatively we will take $\sigma_I = 0.1\arcsec$ for all images,
assuming that their positions are measured on HST images.

$\chi^2$ computed from Eq.(\ref{chi2}) in the source plane is
mathematically equivalent to $\chi^2$ computed in the image plane,
written as:
\begin{equation}
 \chi^2=\displaystyle{\sum_{i=1}^n \sum_{j=1}^{n^i}
\frac{(\vec{\theta_{I_j^i}}-\vec{\theta_{IG_j^i}})^2}{\sigma_{I_j^i}^2}},
\end{equation}
 
\noindent where $\vec{\theta_{IG_j^i}}$ is the image of $\vec{\theta_{SG^i}}$
close to $\vec{\theta_{I_j^i}}$. Indeed
$\vec{\theta_{S_j^i}}-\vec{\theta_{SG^i}}\equiv\vec{\delta S^i_j}$ and
$\vec{\theta_{I_j^i}}-\vec{\theta_{IG_j^i}}\equiv\vec{\delta I^i_j}$
are assumed to be small quantities compared to the variation scale of
the elements of the magnification matrix $\mathcal A_j^i$. Therefore
the local transformation from the image plane to the source plane is
written as $\vec{\delta I^i_j}=\mathcal A_j^i\,\vec{\delta S^i_j}$.
The main motivation for working in the source plane is numerical simplicity
because the
mapping from the source to the image plane is not a one-to-one mapping
and we may not recover all the images when solving the lens equation.

If $M_p$ is the number of fitted parameters for the potential, there
is a total of $M=M_p+2$ adjustable parameters (including $\Omega_m$
and $\Omega_\lambda$) and $N_C$ independant data points. We compute a
$\chi^2$-distribution for $\nu=N_C-M$ degrees of freedom. In practice,
in our simulation we try to recover only the most important parameters, 
like $\sigma_0$ (or $\sigma_c$), $\theta_a$ or $\theta_s$, to limit the 
number of degrees of freedom. This would be the case in a real application.

\subsection{Numerical simulations in different configurations}
\begin{figure}
  \resizebox{\hsize}{!}{\includegraphics{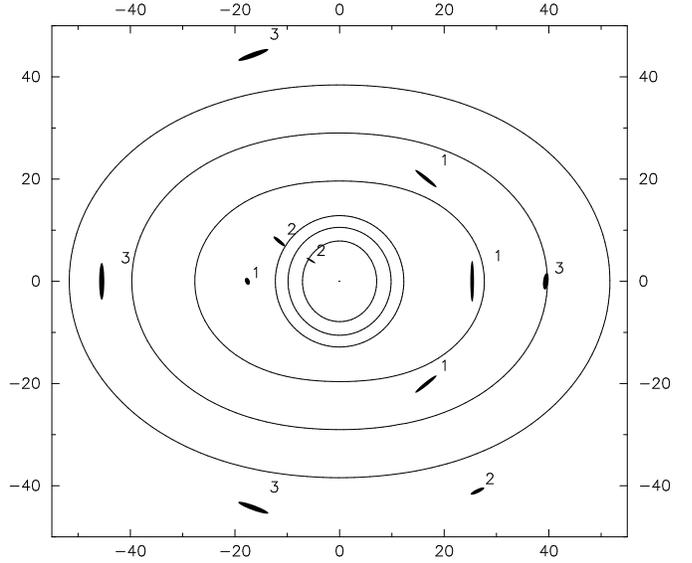}}
\caption{
  Multiple images generated by a pseudo-elliptical NFW cluster at 
  $z_\mathrm{L}=0.3$ with the lens
  parameters: $v_c=2000$ km/s, $\theta_s=31.3$\arcsec 
  ($r_s=150$~kpc)
  and $\epsilon=0.1$. Close to their respective
  critical lines, 3 families of multiple images are identified: a
  tangential one (\# 1, $z_{S1}=0.6$), a radial one (\# 2, $z_{S2}=1$)
  and another tangential one (\# 3, $z_{S3}=4$).
  Units are given in arcseconds.}
\label{3fam}
\end{figure}

To recover the most important parameters of the potential, we
generated 3 families of multiple images (2 tangential ones and a
radial one for a total number of constraints $\nu=16$, see
Fig.~\ref{3fam} and Table \ref{tablez}) with the pseudo-elliptical 
NFW profile developed in \citet{GolseNFW}. We also chose
regularly distributed source redshifts (Table~\ref{tablez}). The 4
geometrical parameters of the cluster lens were left fixed during the
minimization $x_0=y_0=0$, $\theta=0 \degr$ and $\epsilon=0.1$), while
the 2 parameters of the potential ($v_c$ and $\theta_s$) were
allowed to vary.  The initial values for these parameters, used to
create the set of images, correspond to reasonable values found in
cluster lenses: $\theta_s = 31.3\arcsec$ (i.e. 150~kpc) and $v_c
= 2000$ km~s$^{-1}$. This last value corresponds to a
``classical'' central velocity dispersion $\sigma_0=1230$~km~s$^{-1}$
for a HK model (see
Sect.~\ref{change_profile}).  $(\Omega_m^0,\Omega_\lambda^0)$ were
fixed to the $\Lambda$CDM values $(0.3,0.7)$.

\begin{table}
\caption[]{Details on the 3 sets of multiple images used in the
  simulations in Sect.~\ref{simple} and \ref{change_profile}. $n_i$ 
  represents the number of images used for each
  family. It does not include the central de-magnified image created for
  tangential images. $N_C$ is the number of constraints in the lens 
  modeling for each family. $N_C=2\times n_i-2$ for $x$ and $y$ position. The
  unknown position of the source $(x_S,y_S)$ is then removed, reducing $N_C$
  by 2 units.}
\label{tablez}
\begin{flushleft}
\begin{tabular}{lccll}
\hline\noalign{\smallskip}
Family & Type & $n_i$ & $z_\mathrm{S}$ & $N_C$ \\
\noalign{\smallskip}
\hline\noalign{\smallskip}
$i=1$ & Tangential & 4 & 0.6 & 6 \\
$i=2$ & Radial & 3 & 1. & 4 \\
$i=3$ & Tangential & 4 & 4. & 6 \\
\noalign{\smallskip}
\hline
\end{tabular}
\end{flushleft}
\end{table}

\subsubsection{Simple cluster potential} 
\label{simple}
\begin{figure}
  \resizebox{\hsize}{!}{\includegraphics{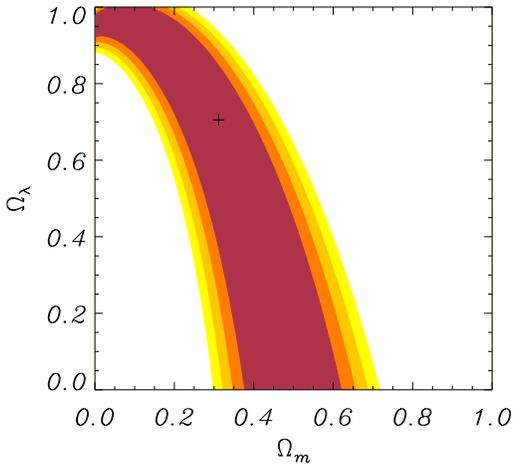}}
\caption{
  $\chi^2$ confidence levels in the $(\Omega_m,\Omega_\lambda)$ plane
  obtained from the optimisation of the lens configuration shown in
  Fig.~\ref{3fam}. The 2 main cluster parameters $v_c$ and
  $\theta_s$ were recovered with $\chi^2_\mathrm{min}=0$. The cross
  (+) represents the original values
  $(\Omega_m^0,\Omega_\lambda^0)=(0.3,0.7)$. 
  Dark to light colors delimit the confidence levels (from 1-$\sigma$
  to 4-$\sigma$).}
\label{Chi2_3fam}
\end{figure}

\begin{figure}
  \resizebox{\hsize}{!}{\includegraphics{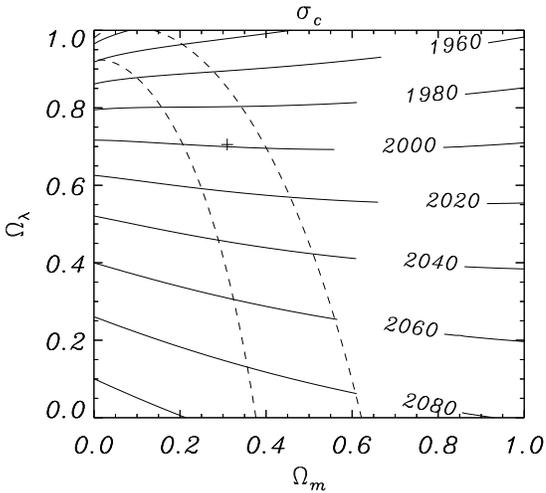}}
\caption{
  Solid lines: distribution of the best-fit velocity dispersion
  $v_c$ from the optimisation of the lens configuration shown in
  Fig.~\ref{3fam}, for each cosmological model.  The cross (+)
  represents the original value for
  $(\Omega_m^0,\Omega_\lambda^0)=(0.3,0.7)$: $v_c=2000$ km
  s$^{-1}$.  Dashed lines correspond to the 1$\sigma$ confidence level
  contours from Fig.~\ref{Chi2_3fam}.  }
\label{sigma3fam}
\end{figure}

In this case, the number of degrees of freedom is $\nu = 16 - 4 = 12$
as 2 cluster parameters are fitted.  The confidence levels of the
minimization are plotted in Fig.~\ref{Chi2_3fam}. The trajectory of
the minimum includes the initial point $(0.3,0.7)$ in the
$(\Omega_m,\Omega_\lambda)$ plane with $\chi^2_\mathrm{min}=0$. The
degeneracy in the cosmological parameters is found as expected in
Fig.~\ref{E2omla}. Tighter constraints can be deduced on $\Omega_m$
than on $\Omega_\lambda$. We also recover the cluster parameters quite
satisfactorily with: $v_c=2000^{+90}_{-90}$ km s$^{-1}$ 
(Fig.~\ref{sigma3fam}) and $\theta_a=31.3 \arcsec ^{+1.2}_{-1.3}$. 
Note that these errors represent only the variations of
the fitted parameters when we scan the $(\Omega_m,\Omega_\lambda)$
plane during the optimisation process.

This preliminary step corresponds to the ``ideal'' case where we
recover the same type of potential we used to generate the images.
Moreover, the morphology of the cluster is regular without
substructure, and we included one radial system among the families of
multiple images. These images are known to probe the
cluster core efficiently. Finally, the redshift distribution of the sources is
wide and the selected redshifts are well separated, for an optimal
sampling of the E-term. One could ask whether any such lens
configuration has already been detected among the known cluster
lenses. It seems that the case of MS2137.3--2353 ($z_\mathrm{L} = 0.31$) is quite
close to this type of configuration \citep{Mellier2} with at least 3
families of multiple images, including a radial one. Uunfortunately, no
spectroscopic redshift has been determined for any of the images so far.

\subsubsection{Changing the shape of the mass profile}
\label{change_profile}
\begin{figure}
  \resizebox{\hsize}{!}{\includegraphics{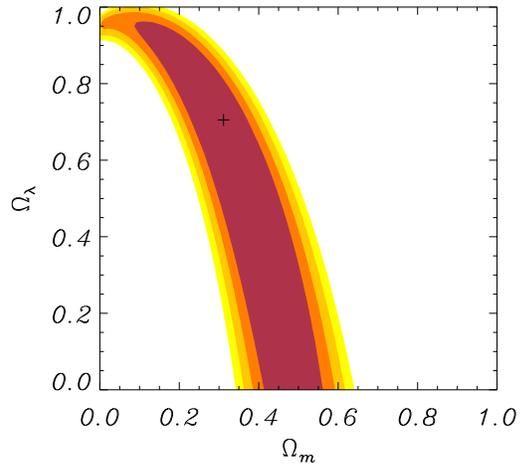}}
\caption{
  $\chi^2$ confidence levels in the $(\Omega_m,\Omega_\lambda)$ plane
  obtained from the optimisation of the lens configuration shown in
  Fig.~\ref{3fam}. In this plot, the potential was fitted with a model
  different from the initial one (an elliptical HK profile instead of
  a pseudo-elliptical NFW). 
  Dark to light colors delimit the confidence levels (from 1-$\sigma$
  to 4-$\sigma$).}
\label{Chi2_3fam_exp}
\end{figure}

To test the sensitivity of the method to the chosen fiducial mass
profile, we tried to recover the lens with another potential, namely
an elliptical HK profile, keeping the same simulated lens.
$\sigma_0$, $\theta_a$ and $\theta_s$ were left free for the
optimization. We first optimized the geometrical parameters for an
arbitrary choice of cosmological parameters.  The best values found
are: $x_0=0.059\arcsec$, $y_0=0.063\arcsec$, $\theta=-0.063\degr$, and
$\epsilon=0.280$. These values are close to the generating ones
($x_0=y_0=0\arcsec$, $\theta=0\degr$), except for the ellipticity
which does not correspond to the same physical meaning in the
pseudo-elliptical NFW profile \citep{GolseNFW}. They were then kept fixed
for the rest of the optimization.  For the lens parameters, we found
$\sigma_0=1230^{+50}_{-50}$ km s$^{-1}$,
$\theta_a=4.6\arcsec^{+0.2}_{-0.1}$ and
$\theta_s=190\arcsec^{+20}_{-10}$. The confidence levels in the
$(\Omega_m,\Omega_\lambda)$ plane are displayed in
Fig.~\ref{Chi2_3fam_exp}.

Although the reconstruction with a potential model different from the
initial ``real'' one does not perfectly fit the data, the results are
quite satisfactory. The confidence levels are even tighter than in the
previous case, but the HK-type potential is characterised by one
additional parameter or equivalently one degree of freedom less ($\nu
= 11$), compared to the pseudo-elliptical NFW profile. Nevertheless we
find a minimum reduced $\chi^2=5$ rather far from 0.

Several other mass profiles were tested as we wanted to discriminate
between the different families of density profiles and test their
sensitivity in the estimate of the cosmological parameters after the
lens reconstruction. We used 5 types of profiles, namely: \\
i) the pseudo-elliptical NFW profile
\citep{GolseNFW}, \\
ii) the singular isothermal ellipsoid (SIE) with $\rho(R)=\rho_0
/ R^2$, $R$ being the elliptical coordinate (Eq.(\ref{R})), \\
iii) the isothermal ellipsoid with core radius (CIE), obtained by
replacing $R$ by $\sqrt{R^2+a^2}$ in the previous expression
\citep[see][] {Kovner}, \\
iv) the HK profile (Eq.(\ref{hjorthkneib})), \\
v) and the King profile characterised by
\begin{equation}
\rho(R)=\rho_0\ \frac{1+
\frac{1-2\alpha}{3} \, R^2 / a^2}
{\left(1+R^2/a^2 \right)^{2+\alpha}}.
\end{equation}

The first 2 profiles are cusped, while the latter have a core
radius and then an additional parameter. For each mass model, we
generated the system of images defined in Table \ref{tablez} (except for
the SIE for which the radial system consists only of 2 images). We
then fitted these images with the other 4 models. All the lens
parameters were left free in this optimisation to get the minimum
reduced $\chi^2$. We did not change the cosmological parameters in
these recoveries. The results are presented in Table
\ref{table_profiles}. We note that the ``core-radius'' profiles
(especially the HK and King ones) can easlily recover the systems
generated by any other models. Indeed in the fit of cusped lens images
by shallower profiles, the core radius can be reduced to very small
values to mimic a large density slope near the center. This is not the
case for the cusped models which cannot mimic images given by a finite
core radius lens model.

\begin{table}
\caption[]{Results of the lens reconstruction using a mass  model
  different from the one used to generate the systems of images. The
  minimum reduced $\chi^2$ is given for each simulation. 
}
\label{table_profiles}
\begin{flushleft}
\begin{tabular}{lccccc}
\hline\noalign{\smallskip}
 Input profile & HK & King & CIE & NFW & SIE \\
\noalign{\smallskip}
\hline\noalign{\smallskip}
Fitted profile & & & & & \\
HK ($\nu=11$) & 0. & 23. & 72. & 460. & 4500. \\
King ($\nu=11$) & 33. & 0. & 33. &  150. & 1500. \\
CIE ($\nu=11$) & 23. & 0.26 & 0. & 87. & 2800. \\
NFW ($\nu=12$) & 6.2 & 21. & 18. & 0. & 680. \\
SIE ($\nu=12$) & 0.14 & 0.011 & 0.28 & 76. & 0. \\
\noalign{\smallskip}
\hline
\end{tabular}
\end{flushleft}
\end{table}

\subsubsection{Influence of the number of multiple systems}
\label{number_syst}

\begin{figure*}
  \resizebox{\hsize}{!}{
\includegraphics{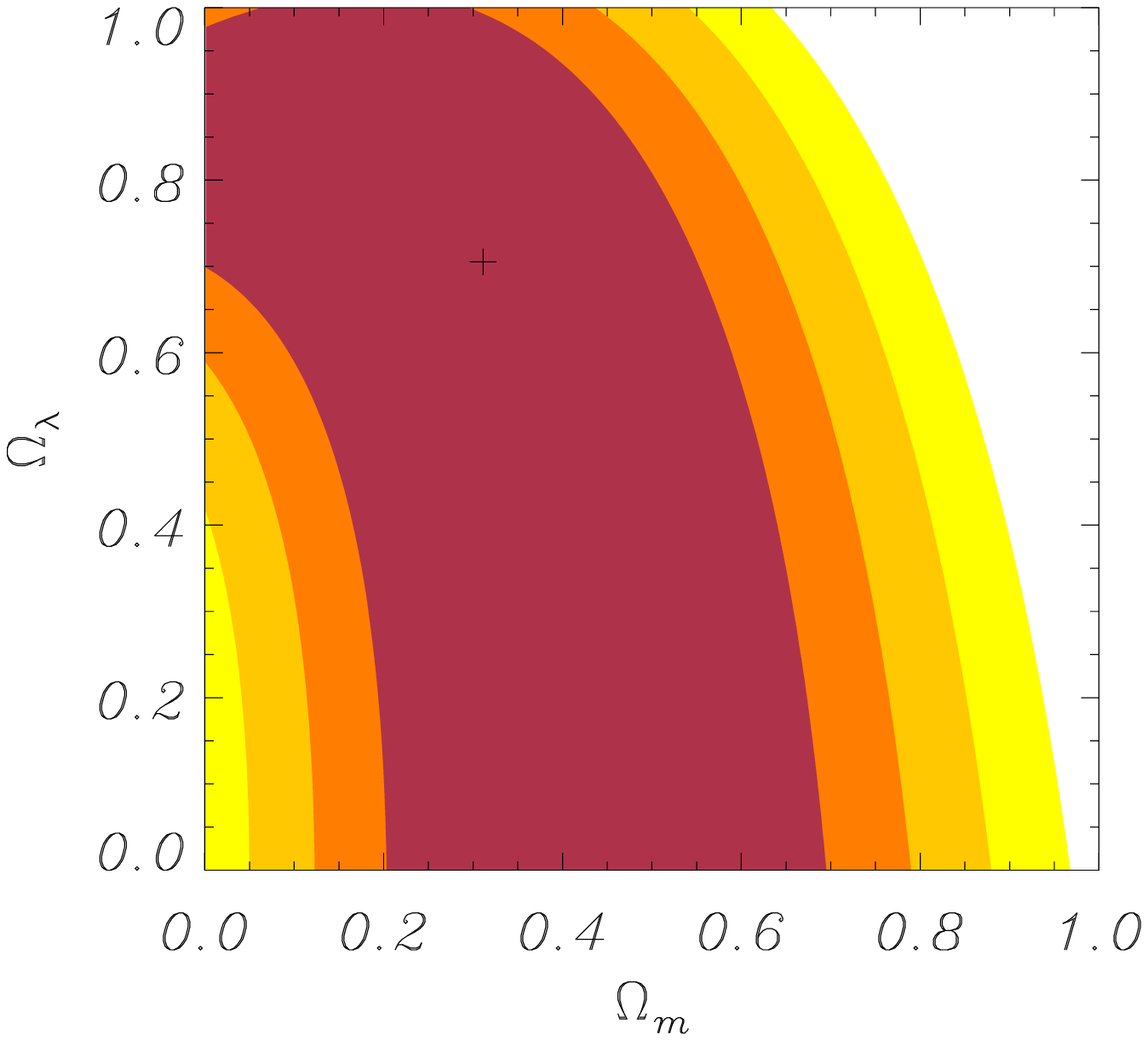}
\hspace*{-3cm}
\includegraphics{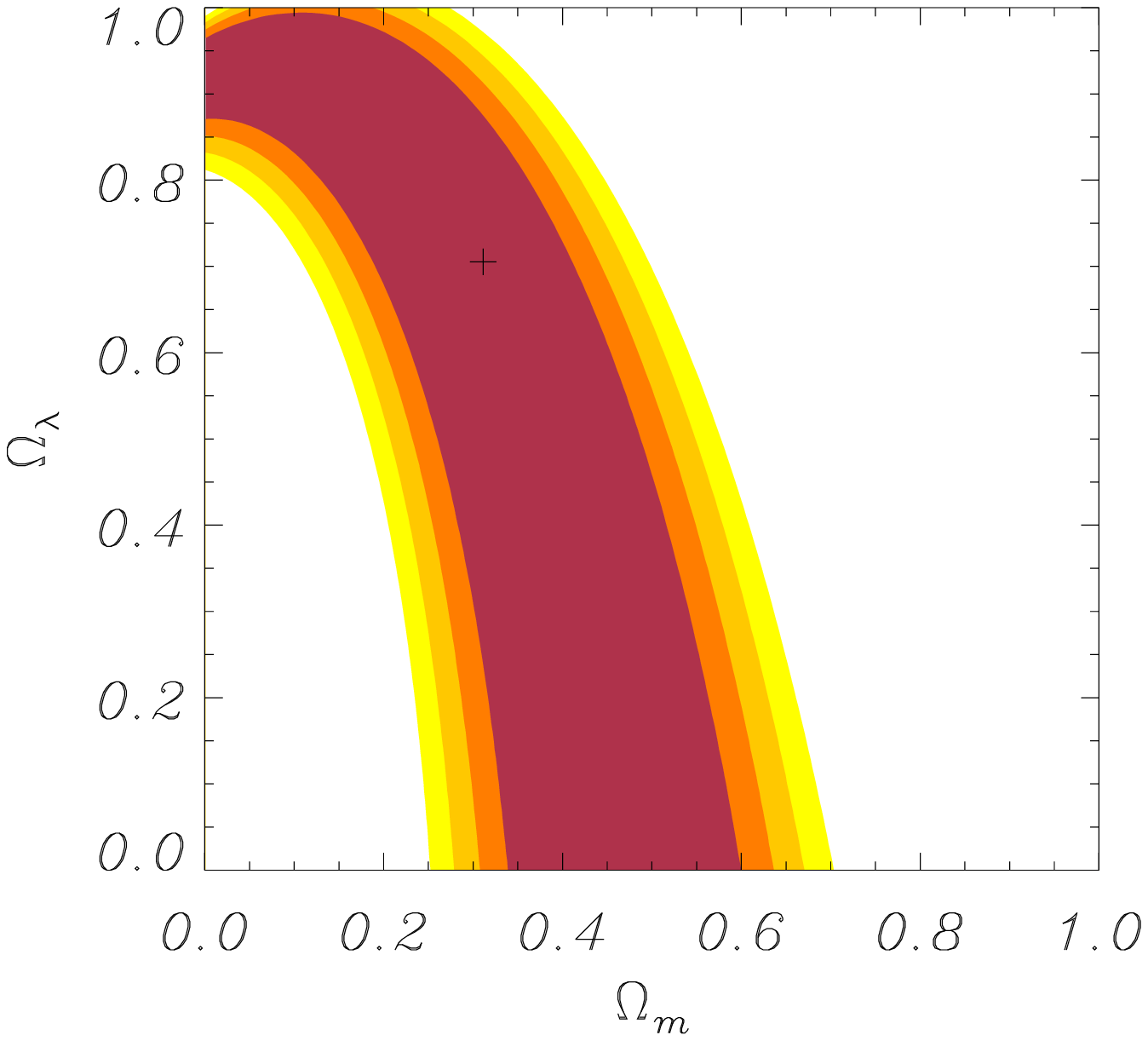}
\hspace*{-3cm}
\includegraphics{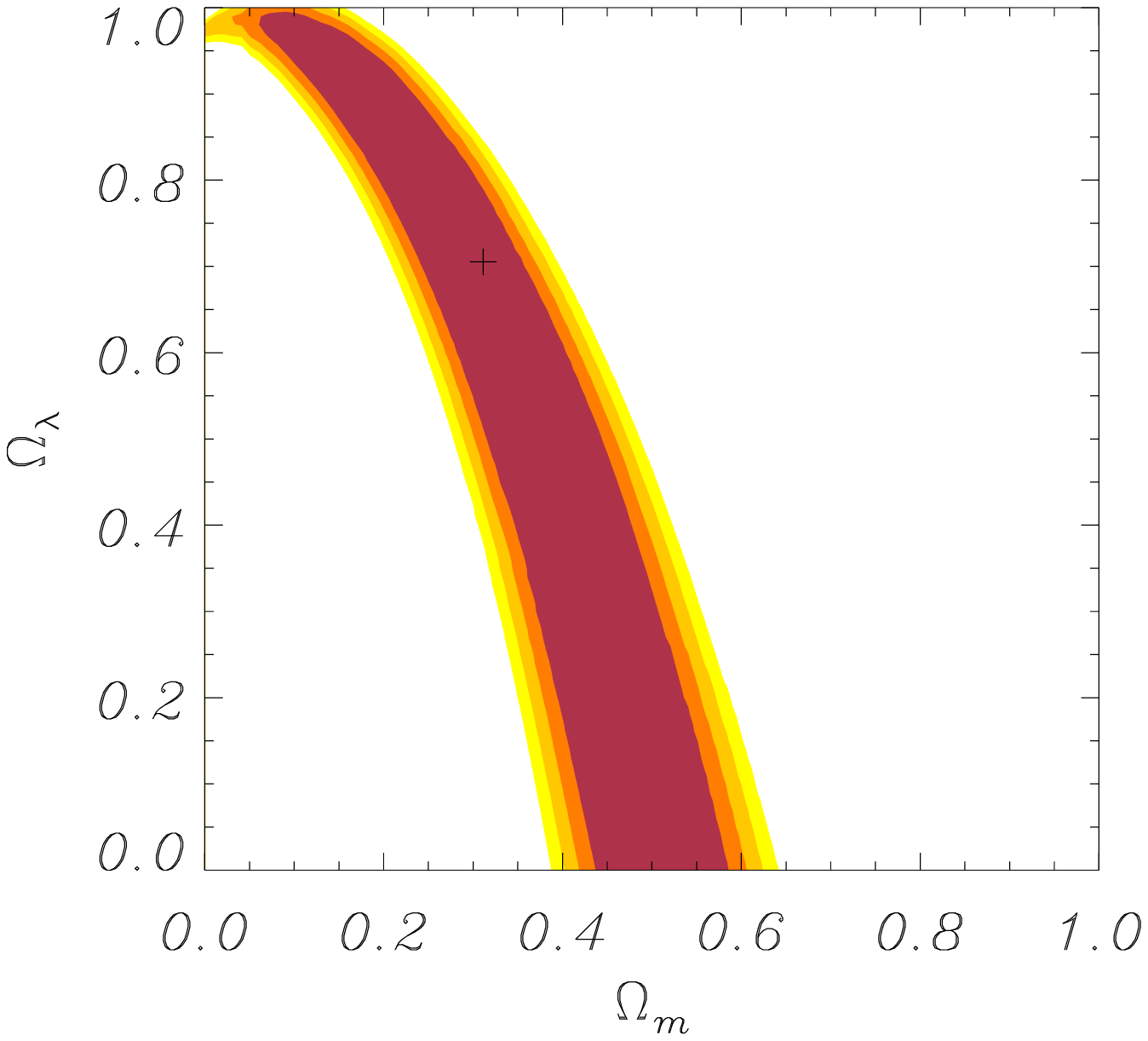}
}
\caption{
  $\chi^2$ confidence levels in the $(\Omega_m,\Omega_\lambda)$ plane
  obtained from the optimisation of the lens configuration described
  in Table \ref{config}. {\bf Left:} 2 systems and $\nu = 10-4=6$
  degrees of freedom. {\bf Middle:} 3 systems and $\nu = 16-5=11$
  degrees of freedom. {\bf Right:} 4 systems and $\nu = 20-5=15$
  degrees of freedom.  The cross (+) represents the original values
  $(\Omega_m^0,\Omega_\lambda^0)=(0.3,0.7)$.
  Dark to light colors delimit the confidence levels (from 1-$\sigma$
  to 4-$\sigma$).}
\label{Chi2_234}
\end{figure*}

In the preceding sections we considered 3 systems of multiple images. As
the method proposed is based on the difference of angular distance
ratios for different redshift planes, we now investigate the influence
of the number of image families. The potential model is again an
HK-type profile at $z_\mathrm{L}=0.3$ with $\sigma_0= 1400$ km s$^{-1}$,
$\theta_a=13.5$\arcsec (i.e. 65 kpc), $\theta_s=146$ \arcsec (i.e. 700
kpc) and $\epsilon=0.2$.  With 2 systems of images, we consider only 2
free parameters for the cluster, because there are not enough
observables to yield results for more parameters, while in the other
cases, 3 parameters are fitted. In all cases, these parameters are
strongly constrained by the fit. Table \ref{erreurs} reports the 
errors on the fitted parameters in the optimisation process, for the different
sets of multiple images detailed in Table \ref{config}.  
The differences in the fitted parameters between the different cases
are small, as they are already well constrained with a single multiple images 
system.

\begin{table}
\caption[]{Recovering of the free parameters of the lens potential for
the Table \ref{config} different systems of images. The errors represent the 
variation of
the fitted parameters at 1-$\sigma$ level when scanning the 
$(\Omega_m,\Omega_\lambda)$ plane in the optimisation process.
 }
\label{erreurs}
\begin{flushleft}
\begin{tabular}{clll}
\hline\noalign{\smallskip}
Nb of systems & $\sigma_0$ (km s$^{-1}$) & $\theta_a$ (\arcsec) & 
$\theta_s$ (\arcsec) \\
\noalign{\smallskip}
\hline\noalign{\smallskip}
2 & $1400^{+60}_{-60}$ & $13.5^{+0.25}_{-0.15}$ & \ \ \ -- \\
\noalign{\smallskip}
\hline\noalign{\smallskip}
3 & $1400^{+70}_{-60}$ & $13.5^{+0.3}_{-0.2}$ & $146^{+2}_{-2}$   \\
\noalign{\smallskip}
\hline\noalign{\smallskip}
4 & $1400^{+60}_{-60}$ & $13.5^{+0.3}_{-0.2}$ & $146^{+14}_{-6}$   \\
\noalign{\smallskip}
\hline
\end{tabular}
\end{flushleft}
\end{table}

The expected constraints on $(\Omega_m, \Omega_\lambda)$ tighten when
the number of families of multiple images increases
(Fig.~\ref{Chi2_234}), especially when their redshift distribution is
wide. 2 families would only provide marginal information on the
cosmological parameters whereas 4 spectroscopically measured systems
would give very tight error bars, provided they are well
distributed in redshift.

\begin{table}
\caption[]{Sets of multiple images used in the  simulations to test
  the influence of their number. $n_i$ represents the number of 
  images used for each
  family. It does not include the central de-magnified image created for
  tangential images. $N_C$ is the number of constraints in the lens 
  modeling for each family. $N_C=2\times n_i-2$. }
\label{config}
\begin{flushleft}
\begin{tabular}{ccccll}
\hline\noalign{\smallskip}
Nb of systems & Family & Type & $n_i$ & $z_\mathrm{S}$ & $N_C$ \\
\noalign{\smallskip}
\hline\noalign{\smallskip}
2 & $i=1$ & Tangential & 4 & 0.6 & 6 \\
  & $i=2$ & Radial & 3 & 1. & 4 \\
\noalign{\smallskip}
\hline\noalign{\smallskip}
  & $i=1$ & Tangential & 4 & 0.6 & 6 \\
3 & $i=2$ & Radial & 3 & 1. & 4 \\
  & $i=3$ & Tangential & 4 & 2. & 6 \\
\noalign{\smallskip}
\hline\noalign{\smallskip}
  & $i=1$ & Tangential & 4 & 0.6 & 6 \\
4 & $i=2$ & Radial & 3 & 1. & 4 \\
  & $i=3$ & Tangential & 4 & 2. & 6 \\
  & $i=4$ & Radial & 3 & 4. & 4 \\
\noalign{\smallskip}
\hline
\end{tabular}
\end{flushleft}
\end{table}

\subsubsection{Influence of additional galaxy masses}
\begin{figure}
  \resizebox{\hsize}{!}{\includegraphics{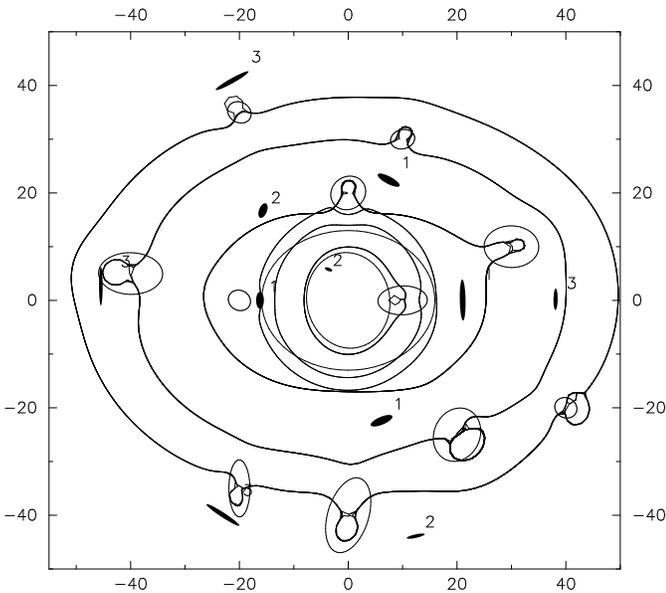}}
\caption{
  Multiple images generated by a cluster at $z_\mathrm{L}=0.3$ with an
  elliptical HK lens profile and the parameters: $\sigma_0=1400$ km/s,
  $\theta_a=13.54$\arcsec ($a=65$~kpc) and $\theta_s=145.8$\arcsec
  ($s=700$~kpc). 12 individual galaxies are added in the potential. 3
  families of multiple images are identified (see Table~\ref{tablez}
  for details). We represent the radial (inside) and tangential 
  (outside) critical lines corresponding to the multiple images redshifts.
  Their characteristic radii are increasing with redshift.
  Units are given in arcseconds.}
\label{Gal}
\end{figure}

In the previous parts, we considered only a main cluster potential
with a regular morphology. We now test the contribution of individual
galaxies, following the prescription used by \citet{Natarajan} as in
Section \ref{mult_sys}. We generated 3 systems of multiple images
formed by the sum of a main cluster with the mass density (HK-type)
characterised by $\sigma_0=1400$ km s$^{-1}$, $\theta_a=13\arcsec$ and
$\theta_s=150\arcsec$ and 12 individual galaxies which represent 30\%
of the total cluster mass (Fig.~\ref{Gal}).

\begin{figure}
  \resizebox{\hsize}{!}{\includegraphics{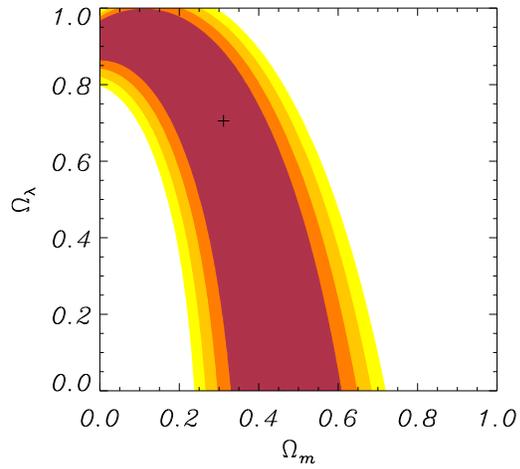}}
\caption{
  $\chi^2$ confidence levels in the $(\Omega_m,\Omega_\lambda)$ plane
  obtained from the optimisation of the lens configuration shown in
  Fig.~\ref{Gal}. For the individual galaxies, we assumed that their
  mass is scaled with the total mass with $\sigma_0^G\propto\sigma_0$.
  The 3 main cluster parameters $\sigma_0$, $\theta_a$ and $\theta_s$
  were recovered with $\chi^2_\mathrm{min}=0$ and $\nu =11$ degrees of
  freedom. 
  Dark to light colors delimit the confidence levels (from 1-$\sigma$
  to 4-$\sigma$).}
\label{Chi2_Gal}
\end{figure}

The images were reconstructed using a main cluster potential with the
same kind of shape as the initial one and the contribution of the
galaxies scaled with $\sigma_0^G$. In addition, we fixed $\sigma_0^G$
proportional to $\sigma_0$ to avoid an increase of the number of free
parameters. Consequently, any variation in $\sigma_0$ means a
rescaling of the total mass of the cluster. So at first order we find
that $\sigma_0^2 \ E$ is constant when we scan the
$(\Omega_m,\Omega_\lambda)$ plane.  Keeping the geometrical parameters
fixed ($x_0=y_0=0\arcsec$, $\theta=0\degr$, and $\epsilon=0.2$), we
obtain the confidence levels in the $(\Omega_m,\Omega_\lambda)$ plane
plotted in
Fig.~\ref{Chi2_Gal} 
and the following constraints on the potential
parameters: $\sigma_0=1400^{+60}_{-65}$ km s$^{-1}$,
$\theta_a=13\arcsec^{+0.3}_{-0.3}$ and $\theta_s=151\arcsec^{+1}_{-1}$.

\begin{figure}
  \resizebox{\hsize}{!}{\includegraphics{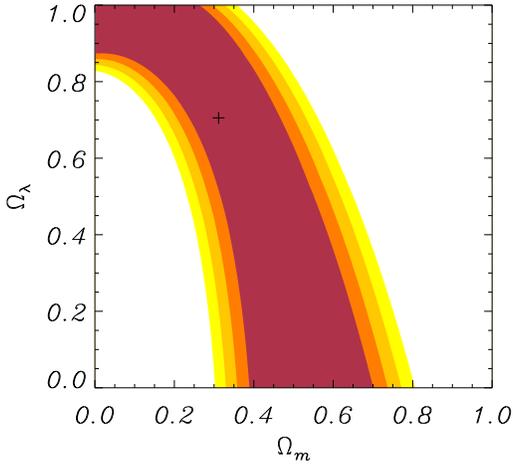}}
\caption{
  $\chi^2$ confidence levels in the $(\Omega_m,\Omega_\lambda)$ plane
  obtained from the optimisation of the lens configuration shown in
  Fig~\ref{Gal}. Here, we did not introduce the individual galaxies
  when recovering the global potential. The 3 main cluster parameters
  $\sigma_0$, $\theta_a$ and $\theta_s$ were recovered but with a
  non-zero reduced $\chi^2_\mathrm{min}$ ($\chi^2_\mathrm{min}=17$).
  Dark to light colors delimit the confidence levels (from 1-$\sigma$
  to 4-$\sigma$).}
\label{Chi2_Gal0}
\end{figure}

To test the influence of the individual galaxies, we tried a
reconstruction without their contribution. For the geometrical
parameters first optimised we obtain $x_0=0.227\arcsec$,
$y_0=0.060\arcsec$, $\theta=-0.748\degr$ and $\epsilon=0.193$, still
close to the generating values. The confidence levels in the
$(\Omega_m,\Omega_\lambda)$ plane are plotted in Fig.~\ref{Chi2_Gal0}.
The contours are slightly shifted and widened compared to the ``good''
ones (Fig.~\ref{Chi2_Gal}) but not significantly different. The
minimum reduced $\chi^2$ is 17. So we are able to correctly retrieve
the cluster potential, even without the individual galaxies
($\sigma_0=1380^{+70}_{-60}$ km s$^{-1}$,
$\theta_a=11.9\arcsec^{+0.3}_{-0.2}$ and
$\theta_s=180\arcsec^{+3}_{-3}$). Adding their contribution is
nevertheless useful to determine precisely the minimum region and to
tighten the confidence levels. It becomes quite critical in more
complex cases or when a single galaxy strongly perturbs the location
of an image.

\subsubsection{Bi-modal cluster mass distribution}
\begin{figure}
  \resizebox{\hsize}{!}{\includegraphics{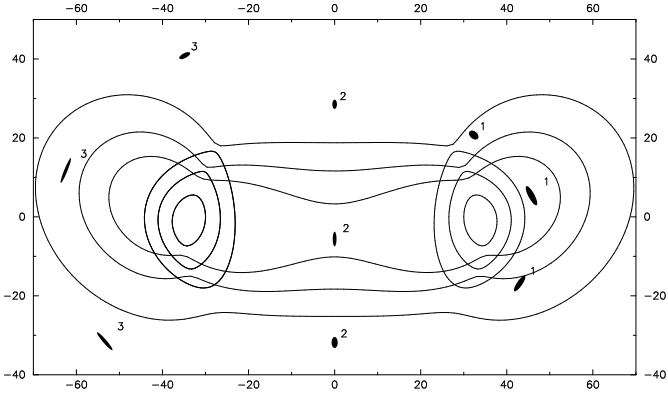}}
\caption{
  Multiple images generated by a bimodal cluster at $z_\mathrm{L}=0.3$ with the
  lens parameters: $\sigma_{01}=\sigma_{02}=1100$ km/s,
  $\theta_{a1}=\theta_a{2}=12\arcsec$ (58~kpc) and
  $\theta_{s1}=\theta_{s2}=167\arcsec$ (800~kpc). 3 families of multiple
  images are identified at $z_{S1}=0.7$, $z_{S2}=1$ and $z_{S3}=2$. 
  Units are given in arcseconds.}
\label{2clumps}
\end{figure}

Up to this point, we have considered simple clusters, dominated by a
single massive component. In reality, most clusters are not fully
virialised and present sub-structure as the result of accretion
processes or merging phases. With these more complex mass
distributions, the lensing configurations are more widely distributed.
Therefore we examine how the cosmological parameters can be
constrained with this type of realistic mass distribution. We thus
generated a bi-modal cluster consisting of two clumps of equal mass
and 3 families of multiple images probing each part of the lens
(Fig.~\ref{2clumps}). The total potential is axisymmetric and each
clump is characterised by an HK-type elliptical mass profile.  As the
number of multiple images is rather small, we limited the number of
parameters to recover and chose $\sigma_0$ and $\theta_a$ for each
clump as adjustable variables. Therefore we fixed $x_{01}=-34\arcsec$,
$x_{02}=34\arcsec$, $y_{01}=y_{02}=0\arcsec$, $\theta_1=-45\degr$,
$\theta_2=+45\degr$, $\epsilon_1=\epsilon_2=0.2$ and
$\theta_{s1}=\theta_{s2}=167\arcsec$.

\begin{figure}
  \resizebox{\hsize}{!}{\includegraphics{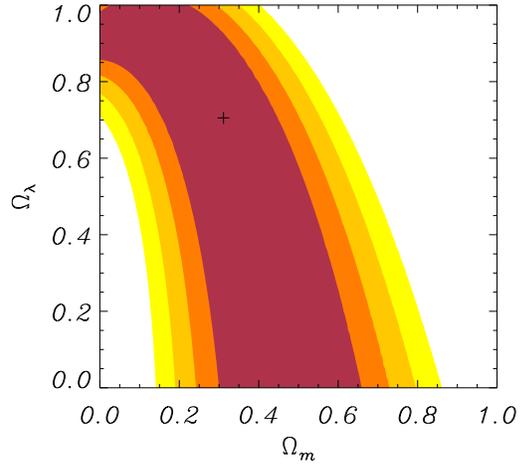}}
\caption{
  $\chi^2$ confidence levels in the $(\Omega_m,\Omega_\lambda)$ plane
  obtained from the optimisation of the lens configuration shown in
  Fig~\ref{2clumps}. The main cluster parameters $\sigma_{01}$,
  $\sigma_{02}$, $\theta_{a1}$ and $\theta_{a2}$ were recovered with
  $\chi^2_\mathrm{min}=0$ and a number of degrees of freedom $\nu =6$.
  The cross (+) represents the original values
  $(\Omega_m^0,\Omega_\lambda^0)=(0.3,0.7)$. 
  Dark to light colors delimit the confidence levels (from 1-$\sigma$
  to 4-$\sigma$).}
\label{Chi2_2clumps}
\end{figure}

Fixing again the initial values of $(\Omega_m^0,\Omega_\lambda^0)$ to
the $\Lambda$CDM model $(0.3,0.7)$, we obtain the confidence levels
plotted in Fig.~\ref{Chi2_2clumps}. The contours are widened compared
to the case of a single potential (in this case, the number of degrees
of freedom is reduced from 11 to 6, but they still give reasonable
constraints). Moreover we note that there is little variation in the
fitted parameters: $\sigma_{01}=1100^{+55}_{-50}$ km s$^{-1}$,
$\sigma_{02}=1100^{+55}_{-45}$ km s$^{-1}$,
$\theta_{a1}=12.1\arcsec^{+0.1}_{-0.1}$, and
$\theta_{a2}=12.1\arcsec^{+0.3}_{-0.2}$. This configuration is close
to the case of the cluster Abell 370, modeled with a bi-modal mass
distribution \citep{Kneib,Bezecourt} needed to reproduce the peculiar
shape of the central multiple-image system.  Unfortunately, up to now
only two redshifts are known for the multiple images identified in A370!

\begin{figure}
  \resizebox{\hsize}{!}{\includegraphics{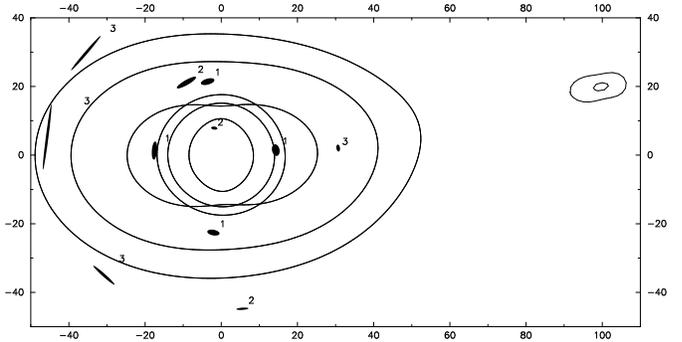}}
\caption{
  Multiple images generated by a cluster at $z_\mathrm{L}=0.3$ consisting of a
  main clump ($\sigma_0=1400$ km/s, $\theta_a=13.54\arcsec$ -- 65~kpc --
  and $\theta_s=145.8\arcsec$ -- 700~kpc) and a smaller one ($\sigma_0=500$
  km/s, $\theta_a=5.2\arcsec$ -- 25~kpc -- and $\theta_s=45.9\arcsec$ --
  220~kpc) located $102\arcsec$ from the main one. Close to their
  respective critical lines, 3 families of multiple images are
  identified: a tangential one (\# 1, $z_{S1}=0.6$), a radial one (\#
  2, $z_{S2}=1$) and another tangential one (\# 3, $z_{S3}=2$).
  Units are given in arcseconds.}
\label{Clump0}
\end{figure}

\begin{figure}
  \resizebox{\hsize}{!}{\includegraphics{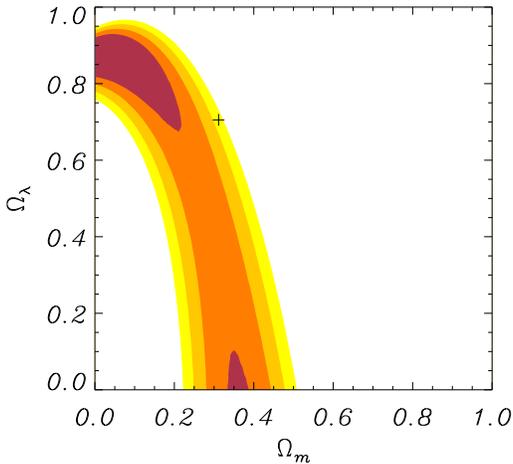}}
\caption{
  $\chi^2$ confidence levels in the $(\Omega_m,\Omega_\lambda)$ plane
  obtained from the optimisation of the lens configuration shown in
  Fig.~\ref{Clump0}. The 3 main cluster parameters $\sigma_0$,
  $\theta_a$ and $\theta_s$ were recovered with a reduced
  $\chi^2_\mathrm{min}=9$ and a number of degrees of freedom $\nu
  =11$. The cross (+) which represents the original values
  $(\Omega_m^0,\Omega_\lambda^0)=(0.3,0.7)$ is now outside the
  3-$\sigma$ confidence levels. 
  Dark to light colors delimit the confidence levels (from 1-$\sigma$
  to 4-$\sigma$).}
\label{Chi2_clump0}
\end{figure}

Last, we generated another system of 3 families of multiple images
produced by a cluster consisting of a main clump ($\sigma_0=1400$ km
s$^{-1}$) and a smaller one ($\sigma_0=500$ km s$^{-1}$) representing
22\% of the total mass (Fig.~\ref{Clump0}). We chose to miss the
small clump in the mass recovery as this may happen when dealing with
some ``dark clumps''. Fitting the configuration with a single main
cluster, we found in a first round the geometrical parameters, which
then remain constant in the $\chi^2$-optimisation: $x_0=0.348\arcsec$,
$y_0=0.189\arcsec$, $\theta=1.880\degr$ and $\epsilon=0.259$. We note in
particular that the ellipticity is larger than the one used to
generate the main clump ($\epsilon=0.2$). This seems to be the
response of the fitting process in order to mimic the missing second
clump.

The parameters left free are again $\sigma_0$, $\theta_a$ and
$\theta_s$. The confidence contours are shown in
Fig.~\ref{Chi2_clump0}. We found the following values of the
parameters: $\sigma_0=1400^{+40}_{-70}$ km s$^{-1}$,
$\theta_a=12.8\arcsec^{+0.2}_{-0.2}$ and
$\theta_s=169\arcsec^{+2}_{-2}$. However in this case, we do not
recover correctly the set of cosmological para\-meters used to
generate the system: $(\Omega_m,\Omega_\lambda)=(0.3,0.7)$ is excluded
at the 3-$\sigma$ level. Moreover the shape of the contours is not the
one expected from the lensing degeneracy. This could be considered to
be a signature of an incorrect fiducial mass distribution due to a
missing clump in the mass reconstruction.  This example demonstrates
that the initial guess and the modeling of the different components
of a cluster are very sensitive elements. They need to be carefully
determined if one wants to test further constraints on the
cosmological parameters


\section{Conclusion and future prospects}

In this paper we have explored in detail a method to obtain
informations on the geometry of the Universe with gravitational
lensing. It follows an approach first presented by Link \& Pierce
(LP98) which states that multiple imaging systems at different
redshifts can provide constraints not only on the mass profile of the
lensing cluster but also on second order parameters like $\Omega_m$ or
$\Omega_\lambda$ -- contained in angular size distances ratios. We
have shown that this technique gives constraints which are degenerate
in the $(\Omega_m, \Omega_\lambda)$ plane and that the degeneracy is
roughly perpendicular to the degeneracy issued from high-redshift 
supernovae searches. Moreover, the matter density $\Omega_m$ can be better
constrained than the $\Lambda$-term. Several simulations of lensing
configurations are proposed, assuming reasonable conditions on
the cluster-lens potential, such as a regular morphology modeled with
only a few parameters.  Provided high quality data can be obtained on
at least 3 systems of multiple images, such as high resolution images
(HST-type) for accurate image positions and deep spectroscopic data
for the measurement of the source redshifts, we can expect
typical error bars of $\Omega_m=0.30{\pm 0.11}$, $\Omega_\lambda=0.70{\pm
  0.23}$.

\begin{table}
\caption[]{List of 6 redshift configurations used in the combination of
  different cluster-lenses (Fig.~\ref{Chi2_combine}) for a global
  $\chi^2$ minimisation. }
\label{table_zs}
\begin{flushleft}
\begin{tabular}{lccc}
\hline\noalign{\smallskip}
$z_\mathrm{L}$ & $z_{S1}$ & $z_{S2}$ & $z_{S3}$ \\
\noalign{\smallskip}
\hline\noalign{\smallskip}
0.15 & 0.4 & 0.8 & 2.0 \\
0.2  & 0.5 & 1.0 & 3.0 \\
0.25 & 0.6 & 0.9 & 2.0 \\
0.3  & 0.6 & 1.0 & 2.0 \\
0.35 & 0.6 & 1.5 & 3.0 \\
0.4  & 0.8 & 1.8 & 4.0 \\
\noalign{\smallskip}
\hline
\end{tabular}
\end{flushleft}
\end{table}

\begin{figure}
  \resizebox{\hsize}{!}{\includegraphics{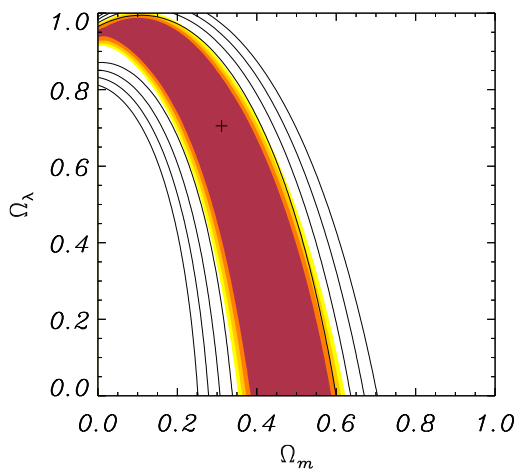}}
\caption{
  \textbf{Color scale}: $\chi^2(\Omega_m,\Omega_\lambda)$ confidence levels
  obtained for a combination of 6 different cluster-lenses
  configurations (see Table~\ref{table_zs} for redshift informations).
  The 3 main cluster parameters $\sigma_0$, $\theta_a$, $\theta_s$
  were recovered for each cluster with a reduced
  $\chi^2_\mathrm{min}=0$ and $\nu=60$ degrees of freedom.  
  \textbf{Dark to light colors delimit the confidence levels (from 1-$\sigma$
  to 4-$\sigma$). Solid} lines: 
  $\chi^2$ confidence levels (from 1-$\sigma$
  to 4-$\sigma$) obtained for a single cluster at
  $z_\mathrm{L} = 0.3$ (same as Fig.~\ref{Chi2_3fam}).  The cross (+)
  represents the original values
  $(\Omega_m^0,\Omega_\lambda^0)=(0.3,0.7)$.  
  }
\label{Chi2_combine}
\end{figure}

It is important to underline that one cluster-lens with adequate
multiple images would provide by itself a strong constraint on the
geometry of the whole Universe.  Such clusters are not that rare:
MS2137.3--2353, MS0440.5+0204, A370, A1689, A2218, AC114 are certainly
good candidates for such an experiment. A thorough and detailed
analysis is still to be done and we have in hand most of the tools to
address the problem immediately. Furthermore, as the exact degeneracy
in the ($\Omega_m$ , $\Omega_\lambda$) plane depends only on the
values of the different redshift planes involved, combining results
from different cluster-lenses can tighten the error bars.  For
illustration, we combined 6 different lens configurations and source
redshifts, as listed in Table~\ref{table_zs}.  Compared to the
expected results with a single cluster (solid lines), the constraints
can be improved significantly (Fig.~\ref{Chi2_combine}).

Looking for a good accuracy on the cosmological parameters is a
permanent search in cosmology. Although the curvature is now
determined with a remarkable precision thanks to recent results from
CMB balloon experiments, it is still very difficult to disentangle
$\Omega_m$ from $\Omega_\lambda$ \citep{Zaldarriaga}. Therefore the
advantages of joint analyses by several independent approaches have
been pointed out (see \citet{White} and \citet{Efstathiou}): combined
results from the $m-z$ relation for SNIa and CMB power spectrum
analyses (which have orthogonal degeneracies) constrain $\Omega_m$ or
$\Omega_\lambda$ separately with much higher accuracy than the
individual experiments alone, leading to the currently favored model. One
impressive example has been given by \citet{Hu} who showed that a
relatively small weak lensing survey could dramatically improve the
accuracy of the cosmological parameters measured by future CMB
missions.The combination of independent tests can improve the
constraints as well as serve as a consistency check. This is clearly
demonstrated by \citet{Helbig0} who combine constraints from lensing
statistics and distant SNIa to get a narrow range of possible values
for $\Omega_\lambda$. Therefore, gravitational lensing is a powerful
complementary method to address the determination of the geometrical
cosmological parameters and probably one of the {\em cheapest} ones,
compared to CMB experiments or SNIa searches.  Our technique, when
applied to about 10 clusters, should be included in such joint
analysis, to obtain a consistent picture on the present cosmological
parameters. We are truly entering an era of accurate cosmology, where
the overlap between the allowed regions of parameter space is
becoming quite reduced.

\begin{acknowledgements}
  We would like to thank Jean-Luc Atteia, Judy Cohen, Harald Ebeling, 
  Richard Ellis,
  Bernard Fort, Yannick Mellier, Peter Schneider and Ian Smail for
  fruitful discussions. We are grateful to Oliver Czoske for a careful reading
  of the manuscript. JPK acknowledges CNRS for support. This work
  benefits from the LENSNET European Gravitational Lensing Network No.
  ER-BFM-RX-CT97-0172.
\end{acknowledgements}

\bibliographystyle{aa}
\bibliography{bibliographie}

\end{document}